\documentclass[fleqn,12pt]{wlscirep}
\usepackage{diagbox}
\usepackage[symbol]{footmisc}
\usepackage{lineno}
\usepackage[sectionbib]{bibunits}

\usepackage[normalem]{ulem}

\usepackage{newunicodechar}
\DeclareRobustCommand{\okina}{%
  \raisebox{\dimexpr\fontcharht\font`A-\height}{%
    \scalebox{0.8}{`}%
  }%
}
\newunicodechar{ʻ}{\okina}

\newcommand{\agn}{J1007\_AGN}

\newcommand{\aap}{Astron. Astrophys.}     
\newcommand\apjs{Astrophys. J. Suppl. Ser.}%
\newcommand\apjl{Astrophys. J}%

\def\Ha     {\ensuremath{\text{H}\alpha}}
\def\Hb     {\ensuremath{\text{H}\beta}}
\def\Hg     {\ensuremath{\text{H}\gamma}}
\def\Hd     {\ensuremath{\text{H}\delta}}
\def\Lya     {\ensuremath{\text{Ly}\alpha}}

\def\neiii     {\ensuremath{\text{[Ne\,\textsc{iii}]}}}

\def\siiv     {\ensuremath{\text{Si\,\textsc{iv}}}}
\def\civ     {\ensuremath{\text{C\,\textsc{iv}}}}
\def\mgii     {\ensuremath{\text{Mg\,\textsc{ii}}}}
\def\heii     {\ensuremath{\text{He\,\textsc{ii}}}}

\def\ciii     {\ensuremath{\text{C\,\textsc{iii]}}}}

\def\heii    {\ensuremath{\text{He\,\textsc{ii}}}}
\def\oiiibr    {\ensuremath{\text{{[O\,\textsc{iii}]}}}}
\def\niv    {\ensuremath{\text{N\,\textsc{iv}}}}
\DeclareMathOperator{\arcsec}{arcsec}

\newcommand{\farcs}{\mbox{$.\!\!^{\prime\prime}$}}%

\def\AVa {$A_V = {2.79}_{-0.25}^{+0.25}\,\textrm{mag}$}
\def\shortAVa {$A_V = {2.79}\,\textrm{mag}$}

\def\fitslope {$\alpha_\lambda = {-1.98}_{-0.27}^{+0.23}$}
\def\fitfsc {$\log_{10}(f_{sc}) = {-1.74}_{-0.21}^{+0.19}$}
\def\fitflux {$f_{2500} =  {21.61}_{-6.58}^{+9.44}$}

\def\AVb {$A_V = {4.47}_{-3.37}^{+1.16}\,\textrm{mag}$}

\def\AVgal {$A_{V,\textrm{gal}} = {0.93}_{-0.18}^{+2.27}\,\textrm{mag}$}

\def\fitslopeb {$\alpha_\lambda = {-1.72}_{-0.85}^{+0.38}$}
\def\fitfluxb {$f_{2500} =  {29.00}_{-25.91}^{+38.45}$}

\def\BHM {$M_{\textrm{BH, GH05, LH}\beta}\!=\!11.52_{-4.63}^{+10.11}\times 10^{7}\,\textrm{M}_{\odot}$}

\def\lEdd {$\lambda_{\textrm{Edd, GH05, LH}\beta}\!=\!0.20_{-0.09}^{+0.13}$}

\def\BHMd {$M_{\textrm{BH, GH05, LH}\beta}\!=\!4.51_{-2.32}^{+5.22}\times 10^{8}\,\textrm{M}_{\odot}$}
\def\shortBHMd {$M_{\textrm{BH, GH05, LH}\beta}\!=\!4.51\times 10^{8}\,\textrm{M}_{\odot}$}

\def\Lbold {$\log_{10}(L_{\textrm{bol, H}\beta}/(\textrm{erg}\,\textrm{s}^{-1}))\!=\! 46.64_{-0.12}^{+0.13}$}
\def\shortLbold {$\log_{10}(L_{\textrm{bol, H}\beta}/(\textrm{erg}\,\textrm{s}^{-1})) = 46.64$}

\def\lEddd {$\lambda_{\textrm{Edd, GH05, LH}\beta}\!=\!0.58_{-0.31}^{+0.62}$}

\def\AlphaOpt {$\alpha_{\textrm{OPT}}=0.28$}

\def \rcrosscorr {$r_{0,\textrm{corr}}^{\textrm{LG}}\approx8.17_{-2.38}^{+2.42}\,\textrm{h}^{-1}\,\textrm{cMpc}$}

\def \rautocorr {$r_0^{\textrm{LL}}\approx19.11_{-8.63}^{+11.49}\,\textrm{h}^{-1}\,\textrm{cMpc}$}

\def \logMmin {$\log_{10}(M_{\textrm{halo, min}}/\textrm{M}_{\odot})= 12.02_{-1.00}^{+0.82}$}

\def \lognhalo {$\log_{10}(n_{\textrm{halo, min}}/\textrm{cGpc}^{-3})=-0.24_{-5.86}^{+3.86}$}

\def\fD {$\log_{10}(f_{\textrm{duty}})\approx3.66_{-3.90}^{+5.89}$}

\def\tQ {$\log_{10}(t_{\textrm{LRD}}/\textrm{yr})\approx12.52_{-3.9}^{+5.9}$}

\title{A Little Red Dot at $\mathbf{z=7.3}$ within a Large Galaxy Overdensity}

\author[1,*]{Jan-Torge Schindler}
\author[2,3]{Joseph F. Hennawi}
\author[4]{Frederick B. Davies}
\author[5,4]{Sarah E. I. Bosman}

\author[6,7]{Ryan Endsley}
\author[7,8]{Feige Wang}
\author[7,8]{Jinyi Yang}

\author[9]{Aaron J.\ Barth}

\author[10,11]{Anna-Christina Eilers}
\author[7]{Xiaohui Fan}
\author[12]{Koki Kakiichi}
\author[13]{Michael Maseda}
\author[2]{Elia Pizzati}
\author[2]{Riccardo Nanni}

\affil[1]{Hamburger Sternwarte, Universität Hamburg, Gojenbergsweg 112, D-21029 Hamburg, Germany}
\affil[2]{Leiden Observatory, Leiden University, P.O. Box 9513, 2300 RA Leiden, The Netherlands}
\affil[3]{Department of Physics, Broida Hall, University of California, Santa Barbara, Santa Barbara, CA 93106-9530, USA}
\affil[4]{Max Planck Institut f\"ur Astronomie, K\"onigstuhl 17, D-69117, Heidelberg, Germany}
\affil[5]{Institute for Theoretical Physics, Heidelberg University, Philosophenweg 12, D-69120, Heidelberg, Germany}
\affil[6]{Department of Astronomy, University of Texas, Austin, TX 78712, USA}
\affil[7]{Steward Observatory, University of Arizona, 933 N Cherry Ave, Tucson, AZ 85721, USA}
\affil[8]{Department of Astronomy, University of Michigan, 1085 S. University Ave., Ann Arbor, MI 48109, USA}
\affil[9]{Department of Physics and Astronomy, 4129 Frederick Reines Hall, University of California, Irvine, CA, 92697-4575, USA}

\affil[10]{Department of Physics, Massachusetts Institute of Technology, Cambridge, MA 02139, USA}
\affil[11]{MIT Kavli Institute for Astrophysics and Space Research, Massachusetts Institute of Technology, Cambridge, MA 02139, USA}
\affil[12]{Cosmic Dawn Center (DAWN), Niels Bohr Institute, University of Copenhagen, Jagtvej 128, DK-2200 København N, Denmark}
\affil[13]{Department of Astronomy, University of Wisconsin-Madison, 475 N. Charter St., Madison, WI 53706, USA}

\affil[*]{jtschindler@hs.uni-hamburg.de}

\begin{abstract}
The nature of "Little Red Dots" and their relation to other forms of accreting supermassive black holes remain an open question. Here we report the discovery of a Little Red Dot at $z=7.3$. It is attenuated by moderate amounts of dust, \shortAVa, with an intrinsic bolometric luminosity of $10^{46.6}\,\textrm{erg}\,\textrm{s}^{-1}$ and a SMBH mass of $5\times10^8\,\textrm{M}_\odot$. 
Most notably, this object is embedded in an overdensity of eight nearby galaxies, allowing us to calculate a spectroscopic estimate of the clustering of galaxies around Little Red Dots. We find a Little Red Dot-galaxy cross-correlation length of $r_0\!=\!8\pm2\,\textrm{h}^{-1}\,\textrm{cMpc}$, comparable to that of $z\!\sim\!6$ UV-luminous quasars.  
The resulting estimate of their minimum dark matter halo mass of $\log_{10}(M_{\textrm{halo, min}}/\textrm{M}_{\odot})= 12.0_{-1.0}^{+0.8}$ indicates that nearly all halos above this mass must host actively accreting SMBHs at $z\approx7$, in strong contrast with the far smaller duty cycle of luminous quasars ($<1\%$).
Our results, taken at face value, motivate a picture in which SMBHs in Little Red Dot phases could serve as the obscured precursors of UV-luminous quasars,
which provides a natural explanation for the short UV-luminous lifetimes inferred from both quasar clustering and quasar proximity zones.
\end{abstract}

\begin{document}


\flushbottom
\maketitle
\thispagestyle{empty} 

\newpage

James Webb Space Telescope (JWST) spectroscopy has confirmed dozens of (type-1) active galactic nuclei (AGN) by detection of a broad ($\textrm{FWHM}>1000\,\textrm{km}\,\textrm{s}^{-1}$) emission-line component to the \Ha\ (or \Hb) line\cite{Harikane2023, Kocevski2023, Maiolino2024b, Uebler2023a, Furtak2024,  Greene2024, Matthee2024, Akins2025}, characteristic of gas motion in the gravitational field of a supermassive black hole (SMBH).
A particularly intriguing subclass of these broad-line AGN appear as compact, red sources in JWST/NIRCAM imaging ($\textrm{F277W}-\textrm{F444W}>1$)\cite{Harikane2023,Matthee2024,Greene2024}.
Broad-line AGN photometrically selected with similar criteria have become known as ``Little Red Dots''\cite{Matthee2024} (LRDs).
Their spectral energy distributions are characterized by a red rest-frame optical slope ($\alpha_{\textrm{OPT}}\gtrsim0$; $f_{\lambda} \propto \lambda^{\alpha}$), often observed in combination with a blue rest-frame UV slope ($\alpha_{\textrm{UV}}\lesssim-1$), resulting in V-shaped continua\cite{Matthee2024, Greene2024} with a minimum around $\sim3500\,\textrm{\AA}$.
These have been interpreted as moderately obscured ($A_{\textrm{V}}{=}1{-}4$) AGN\cite{Harikane2023, Matthee2024, Greene2024} superimposed on a galaxy stellar component or a fraction of unattenuated scattered AGN light.
Alternatively, the spectral features could be explained by a star forming galaxy that hosts an AGN encased in a cloud of extremely dense gas\cite{Inayoshi2025, JiXihan2025}.
To date the underlying physical processes that produce the characteristic LRD spectra are not yet fully understood and highly debated\cite{Setton2024, Kokubo2024, Naidu2025, Juodzbalis2025, Rusakov2025}.
While their intrinsic bolometric luminosities and SMBH masses can rival those of quasars, their number densities\cite{Matthee2024, Greene2024, Kocevski2025, Kokorev2024} ($10^3\textrm{--}10^4\,\textrm{Gpc}^{-3}$) place them factors of 10-100 above the faint-end extrapolations of the $z>5$ quasar luminosity functions\cite{Niida2020, Matsuoka2023}.

The existence of $z\gtrsim6$ quasars\cite{Fan2023} with $M_{\textrm{BH}}>10^9\,\textrm{M}_\odot$ SMBHs  challenges models of SMBH formation.
In the canonical picture, SMBH growth is bounded by the Eddington limit and black holes grow exponentially with the Salpeter timescale\cite{Salpeter1964}. With an average radiative efficiency of $\sim0.1$, this e-folding timescale is $\sim50\,\textrm{Myr}$.
To assemble the massive SMBHs of luminous $z\sim6$ quasars, continuous accretion at the Eddington limit comparable to the Hubble time $t_{\textrm{H}}(z)$ is necessary. 
Hence, the quasar duty cycle, the fraction of cosmic time a galaxy shines as a luminous quasar, $f_{\textrm duty}=t_{\textrm{Q}}/t_{\rm{H}}(z)$ is expected to be around unity. 
%
However, clustering measurements place quasars in massive dark matter halos\cite{Arita2023, Eilers2024, Pizzati2024} ($\log_{10}(M_{\textrm{halo}}/M_\odot)\approx 12.30$) and motivate small UV-luminous duty cycles of $f_{\textrm{duty}}\approx0.1\%$, phases of active growth of $t_{\textrm{Q}}\!\sim\!10^6-10^7\,\textrm{yr}$, exacerbating the challenge to grow their SMBHs from stellar seeds to $M_{\textrm{BH}}>10^9\,\textrm{M}_\odot$ by $z\approx6$.
Radiatively inefficient accretion, which directly results in much faster SMBH growth, or UV-obscured, dust-enshrouded growth phases for the bulk of the SMBH population and thus a much larger intrinsic duty cycle are two proposed solutions to this problem\cite{Hopkins2005, Ricci2017, Davies2019, Satyavolu2023, Endsley2023, Eilers2024}.
If LRDs with intrinsic quasar-like properties are found in similarly massive dark matter halos to quasars, they could belong to the same population and would be appealing candidates for obscured phases of quasar growth, because their expected duty cycles would have to approach unity\cite{Pizzati2025}.
In this article, we present the first spectroscopic LRD-galaxy clustering measurement to determine how LRDs are embedded in the evolving cosmic web of dark matter halos, enabling a direct comparison with high-z UV luminous quasars.

The JWST program GO 2073 builds the foundation to map the morphology of the ionized intergalactic medium around two $z\gtrsim7$ quasars for stringent constraints on SMBH growth.
The immediate goals are to identify galaxies at and beyond the quasars' redshift to study quasar-galaxy clustering and to provide targets for subsequent deep spectroscopic observations to map the quasar ``light-echoes''.
Based on JWST NIRCam pre-imaging in four filter bands (F090W, F115W, F277W, F444W) and ground based LBT/LBC photometry (see Methods), we have selected galaxy candidates for spectroscopic NIRSpec/MSA follow-up in the same cycle.
Among the followed-up galaxy candidates, we identified one broad-line AGN at $z=7.26$, \agn, and eight nearby galaxies at similar redshifts ($z=7.2-7.3$) in the field of quasar J1007+2115\cite{YangJinyi2020} ($z=7.51$)

%
%
 
\begin{figure}[ht!]
\centering
\includegraphics[width=0.9\textwidth]{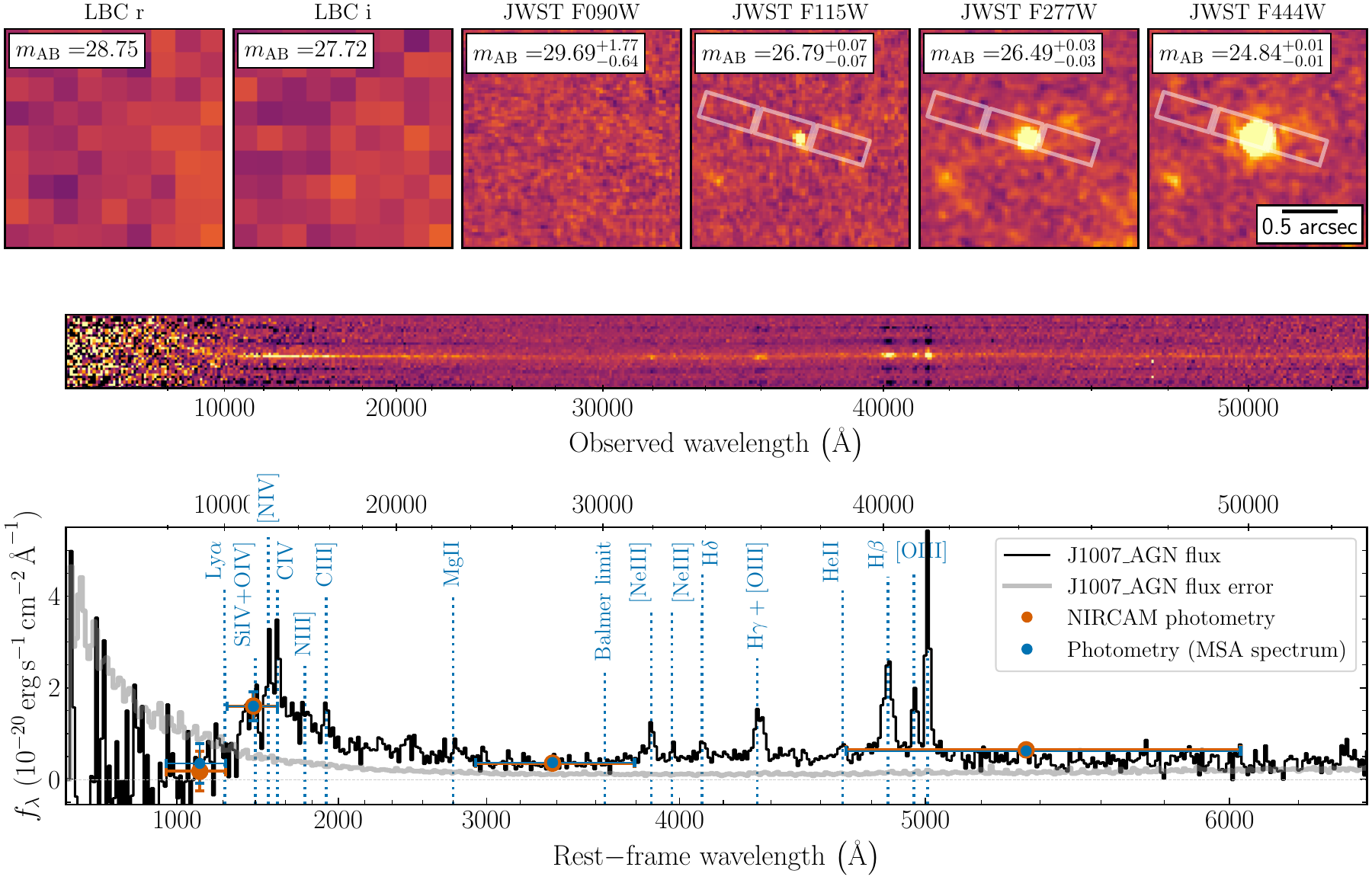}
\caption{\small
\textbf{Photometry and discovery spectroscopy of \agn\ at $\mathbf{z\sim7.3}$.} \textit{Top panel:} Image cutouts ($2\arcsec \times 2\arcsec$) covering the six filters from ground-based LBT/LBC r- and i-band images to JWST/NIRCam F090W, F115W, F277W, and F444W. By design, the source is not detected in the bluest three bands and appears as a red (F115W$-$F444W $=1.94\,\textrm{mag}$) object with compact morphology in F115W, F277W, and F444W. Referenced apparent magnitudes are calculated from aperture photometry at the source location. \textit{Middle panel:} The co-added 2D NIRSpec/MSA spectrum of the three AB subtracted dither positions as a function of observed wavelength. The 2D spectrum is displayed in pixel coordinates, resulting in a non-linear observed wavelength axis. The bright trace is the positive coadded spectrum, whereas the dark traces show the 4 negative traces of the individual AB dithers. 
\textit{Bottom panel:} The 1D optimally extracted co-added spectrum as a function of rest-frame wavelength. Positions of spectral features, including typical line emission observed in AGN, are indicated with blue dotted lines. Orange data points show the fluxes measured from the NIRCam photometry. Errorbars on the photometry denote the $3\sigma$ flux error and the wavelength range of the filter in which the transmission is above 50\% of its peak value. The blue data points show the synthetic photometric calculated from the MSA spectrum. 
}
\label{fig:discovery}
\end{figure}

\begin{figure}[ht!]
\centering
\includegraphics[width=0.55\textwidth]{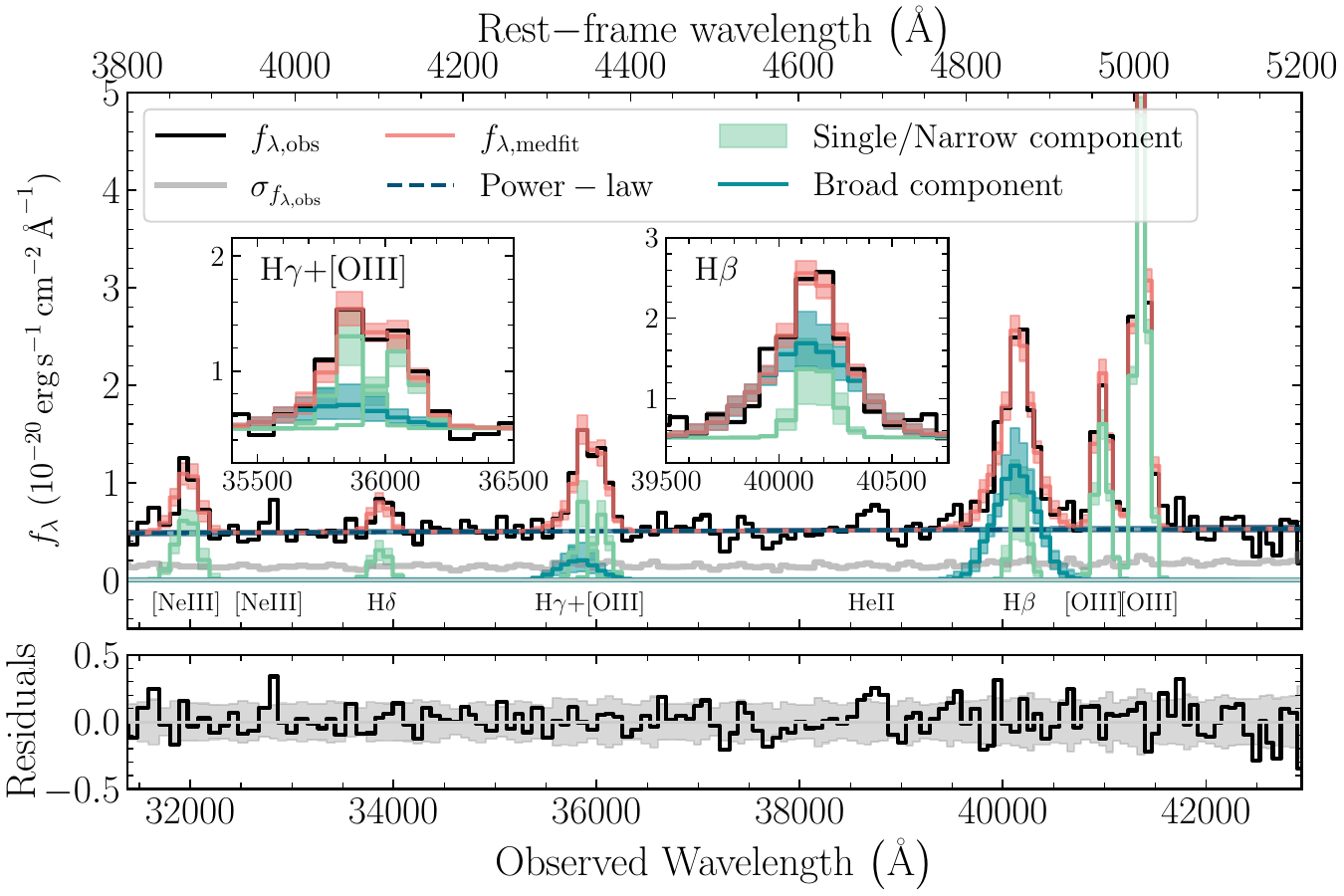}
\includegraphics[width=0.42\textwidth]{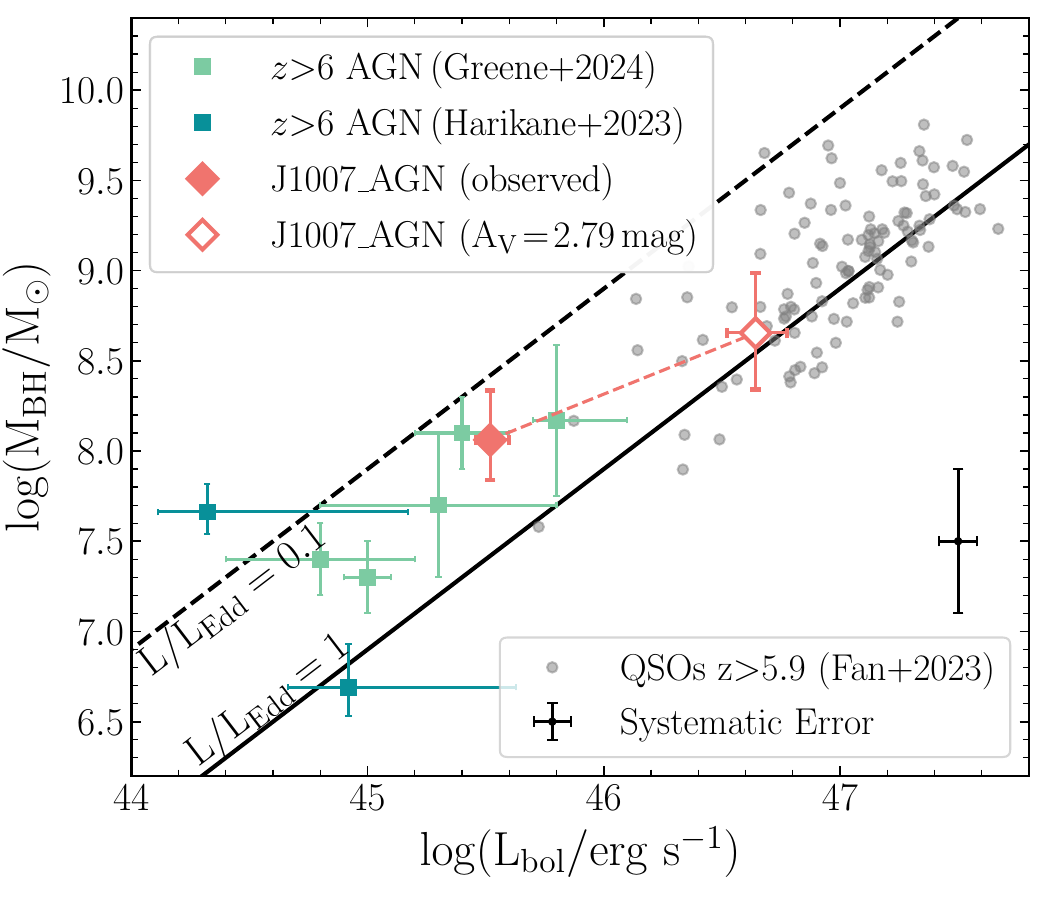}
\caption{\small
\textbf{Spectral modeling and physical properties of \agn.} \textit{Left panel:} 
Posterior median of our model fit (bright red) against the \agn\ spectrum (black). Individual fit components are highlighted with blue and green lines. Light coloured regions depict fit uncertainties (16 to 84 posterior percentile range). Insets illustrate the decomposition of the \Hg\ and \Hb\ emission lines (continuum model added to line components).
\textit{Right panel:} \agn\ (bright red diamonds) in comparison to $z>6$ AGN\cite{Harikane2023, Greene2024} (blue and green squares) and high-redshift quasars from the literature\cite{Fan2023} (gray points, systematic $1\sigma$ uncertainty in black) in the black hole mass bolometric luminosity plane. 
Error bars on the colored data denote statistical $1\sigma$ uncertainties (or  16th to 84th percentile of posterior). 
We differentiate between the observed (filled diamond) and dereddened (\AVa; open diamond) measurements. Uncertainties on $A_V$ are consistently propagated and included in both bolometric luminosity and black hole mass. 
For display, we adopted the fiducial bolometric luminosity $L_{\textrm{bol, H}\beta}$ and the SMBH mass estimate $M_{\textrm{BH, GH05, LH}\beta}$ of Extended Data Tables\,\ref{tab:spec_info} and \ref{tab:dereddened_prop}, which are  based on the \Hb\ emission line properties (see Methods). 
}
\label{fig:agn_analysis}
\end{figure}

Figure\,\ref{fig:discovery} displays the photometric and spectroscopic discovery observations of \agn.
It appears as a compact source with a red rest-frame optical colour (F277W$-$F444W\!=\!$1.65\,\textrm{mag}$; also see Extended Data Figure\,\ref{fig:phot_comp}).
The NIRSpec/MSA PRISM spectrum of \agn\ (Figure\,\ref{fig:discovery}, bottom panel) has a clearly detected continuum, featuring a plethora of strong emission lines.
The most prominent feature is the \oiiibr $\lambda\lambda 4959,5007$ doublet, from which we derive the source redshift of $z=7.2583\pm0.0006$ (see Methods). As shown in Figure\,\ref{fig:agn_analysis} (left), we decompose the spectrum with a multi-component fit (see Methods) and find a broad \Hb\ line width of $\textrm{FWHM}=3370_{-648}^{+1156}\,\textrm{km}\,\textrm{s}^{-1}$ ($R\sim180\approx1650\,\textrm{km}\,\textrm{s}^{-1}$ at \Hb), allowing us to unambiguously classify this source as a type-1 AGN. 
The source also exhibits weak \Hd\ ($\textrm{SNR}\approx3$) and \Hg\ (blended with \oiiibr $\lambda4364$) Balmer lines as well as several high-ionization emission lines (e.g., \neiii\ $\lambda3869.85$; $\textrm{SNR}\approx6$) typical for AGN.
We further detect high-ionisation rest-frame UV lines, notably \niv\ $\lambda1486$ and \civ $\lambda\lambda1548.2,1550.8$ emission (see Methods for details).
Its photometric and spectroscopic properties place this source firmly among the population of LRDs\cite{Matthee2024, Greene2024}.

%
 
%
We measure an absolute magnitude at rest-frame $1450\,\textrm{\AA}$ of $M_{1450}=-19.76_{-0.45}^{+0.77}\,\textrm{mag}$ from the spectrum. 
In comparison to large samples of LRDs\cite{Kocevski2025}, \agn\ stands out as a particularly luminous source.
For the time being, we follow the  literature\cite{Kocevski2023, Greene2024} in interpreting the spectral continuum as a combination of a dust-reddened AGN component, dominating the rest-frame optical, and scattered AGN light, which produces the observed rest-frame UV emission. 
Fitting an appropriate continuum model to the data (see Methods and Extended Data Figure\,\ref{fig:av_fit}), we find \agn\ to be moderately dust obscured with \AVa.

The standard approach to estimate the black hole masses of LRDs utilizes scaling relations\cite{Greene2005} between the line width and the line luminosity of the broad \Hb\ line component.
We carefully decompose the rest-frame optical emission with a parametric model (see Methods) as shown in Figure\,\ref{fig:agn_analysis} (left) and measure the properties of the individual components as summarized in Extended Data Table\,\ref{tab:spec_info}.
Using the measured \Hb\ line properties, we derive a black hole mass of \BHM.
The bolometric luminosity is typically estimated from the rest-frame UV continuum emission. However, in LRDs this emission is not fully understood. Therefore, we convert the \Hb\ line luminosity ($L_{\textrm{H}\beta}$) to a continuum luminosity ($L_{5100}$) using a relation derived from low-z AGN\cite{Greene2005} and then adopt a bolometric correction\cite{ShenYue2011} of $L_{\textrm{bol}} = 9.26 \times L_{5100}$.
We estimate a bolometric luminosity of $\log_{10}(L_{\textrm{bol},\textrm{H}\beta}/(\textrm{erg}\,\textrm{s}^{-1}))=45.52_{-0.06}^{+0.08}$. With an Eddington luminosity ratio of \lEdd, \agn\ is rapidly accreting mass.
These measurements were derived based on the observed spectrum. However, our continuum model suggests that the AGN emission that dominates at the \Hb\ wavelength is attenuated by dust. 
Correcting the spectral model for dust attenuation of \AVa, we derive significantly larger values for the SMBH mass \shortBHMd, bolometric luminosity \shortLbold, and Eddington luminosity ratio \lEddd (also see Extended Data Table\,\ref{tab:dereddened_prop}).
Figure\,\ref{fig:agn_analysis} (right) places these results in the context of quasars\cite{Fan2023} and other $z>6$ AGN\cite{Harikane2023, Greene2024}, using equivalent assumptions to derive bolometric luminosities and black hole masses.
Based on its observed bolometric luminosity and black hole mass, \agn\ straddles  the boundary region between the more luminous LRDs and the faint high-z quasar population\cite{Onoue2019}.
Taking into account the dust attenuation, it could intrinsically be as luminous ($L_{\textrm{bol}}\approx10^{46}\,\textrm{erg}\,\textrm{s}^{-1}$) and massive ($M_{\textrm{BH}}\gtrsim10^{8.5}\,\textrm{M}_\odot$) as a typical UV-luminous quasar. 

\begin{figure}[ht!]
\centering
\includegraphics[width=0.45\textwidth]{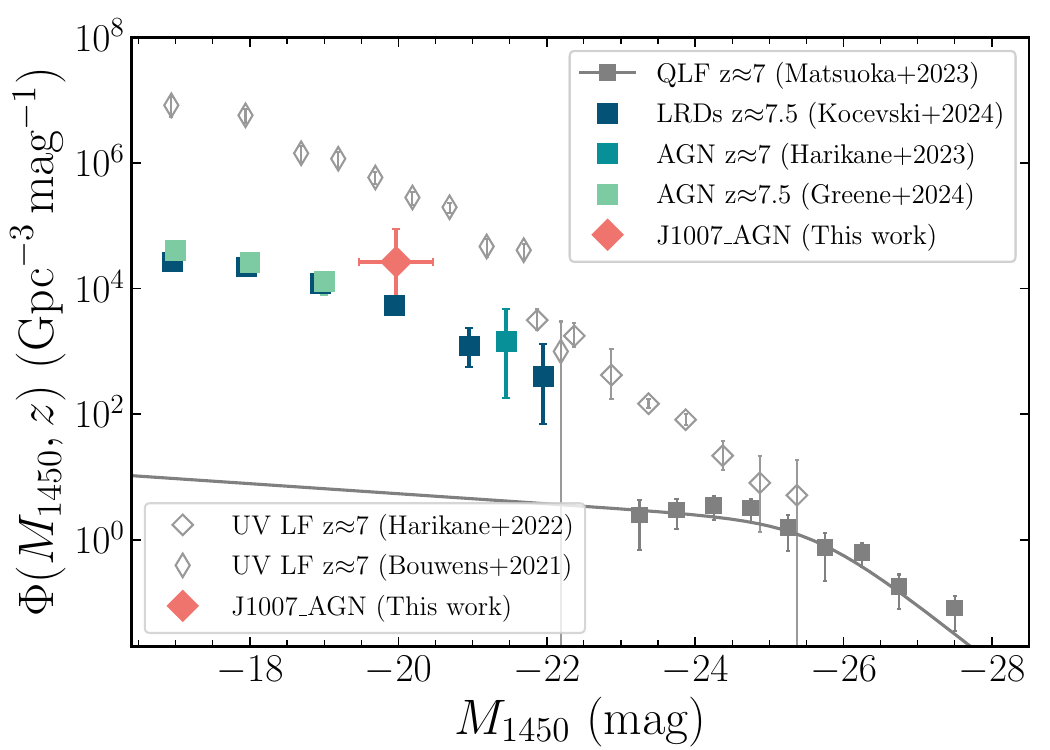}
\includegraphics[width=0.45\textwidth]{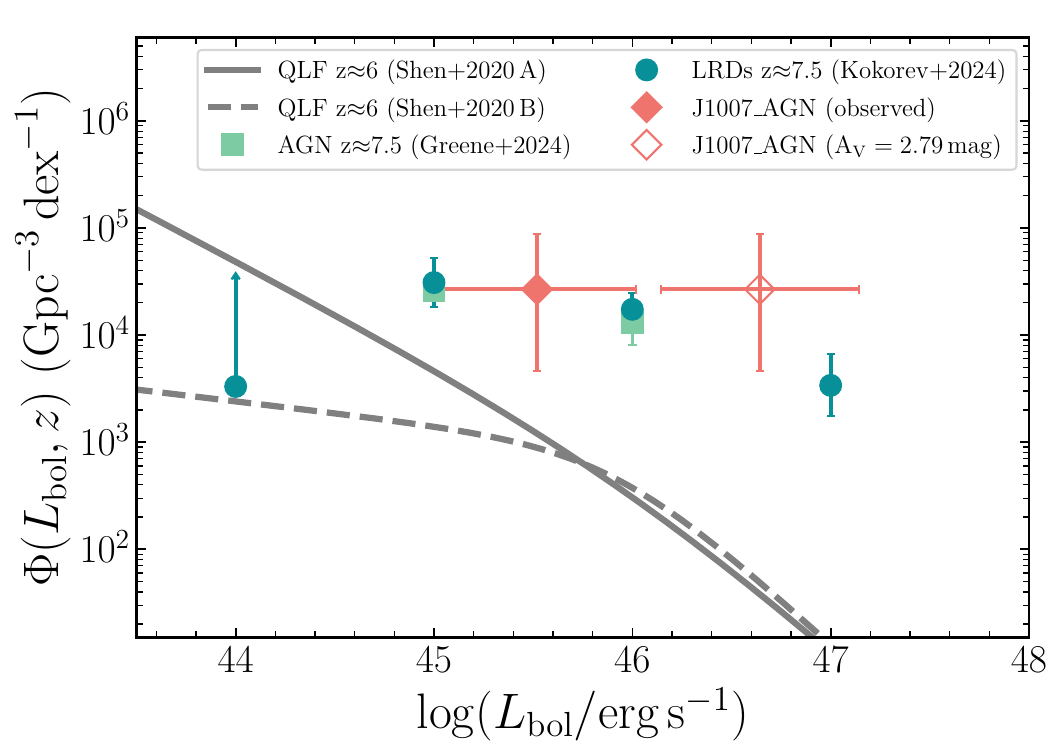}
\caption{\small \textbf{Volume densities of $\mathbf{z>6}$ broad-line AGN and quasars.}
\textit{Left panel:} Binned UV luminosity function estimates of spectroscopically confirmed JWST (red) AGN\cite{Harikane2023, Greene2024} (blue, green squares) and photometrically selected LRDs\cite{Kocevski2025} (dark blue square) in comparison to galaxies \cite{Bouwens2021, Harikane2022} and quasars\cite{Matsuoka2023} at $z\approx7$.
Our estimate based on \agn\ (bright red) agrees well with the other measurements for faint JWST AGN, but falls orders of magnitude above the faint-end extrapolation of the $z\approx7$ quasar luminosity function\cite{Matsuoka2023} (gray solid line).
\textit{Right panel:} Measurements of the bolometric luminosity function for red AGN spectroscopically confirmed with JWST\cite{Greene2024} (green squares) and photometrically selected LRDs\cite{Kokorev2024} (blue circles) in comparison to our estimate (bright red) and two bolometric QLF model fits\cite{ShenXuejian2020} at $z\approx6$ (gray). Model A (solid gray line) has a flexible faint-end slope evolution, whereas the faint-end slope is restricted to evolve monotonically in model B (dashed gray line).
Dereddening \agn\ by \AVa\ increases the bolometric luminosity by a factor of $\sim10$ (open bright red diamond), pushing the source well into the quasar regime, $L_{\textrm{bol}}\gtrsim10^{46}\,\textrm{erg}\,\textrm{s}^{-1}$. 
Vertical error bars on the literature measurements reflect the $1\sigma$ statistical uncertainies. 
For our  luminosity function estimate (bright red) the error bars indicate the $1\sigma$ confidence interval for a $N=1$ Poisson distribution\cite{Gehrels1986}. Error bars in the luminosity direction indicate the luminosity bin width.
}
\label{fig:agn_lf}
\end{figure}

Correcting for our selection function, we estimate a number density of LRDs of $n_{\textrm LRD}\approx2.67\times10^{4}\,\textrm{Gpc}^{-3}$ based on our serendipitous discovery (detailed description in Methods).  
In Figure\,\ref{fig:agn_lf} we compare this number density (orange diamond) with luminosity function measurements for broad-line AGN and LRDs (coloured squares) and quasars (gray filled symbols and lines), as a function of UV magnitude (left) and bolometric luminosity (right). 
The figure underlines that our discovery is consistent with the expectation from $z>6$ LRDs, but $>100$ times above the best constraints of the $z\gtrsim6$ quasar luminosity functions\cite{Matsuoka2023, ShenXuejian2020}.

With the preceding analysis, we established that \agn\ is a bright $z=7.26$ LRD. Exploiting the discovery of eight galaxies in the vicinity of \agn, we conduct a clustering analysis to constrain the environment of a $z\approx7.3$ LRD for the first time. Details on the galaxies can be found in the Methods section (Extended Data Figures\,\ref{fig:galaxy_discovery} and \ref{fig:galaxy_positions}, Extended Data Tables\,\ref{tab:source_info} and \ref{tab:galaxy_info}).
For our fiducial analysis, we restrict ourselves to the six nearest galaxies out of the eight, which are within $\left| \Delta v_{\textrm{LOS}} \right|\!<\!1500\,\textrm{km}\,\textrm{s}^{-1}$. However, the results do not strongly depend on this assumption (see Methods).
Taking our survey selection function and targeting completeness into account, we calculate the volume-averaged LRD-galaxy cross-correlation function $\chi$ in three radial bins (see Methods). 
We find an excess of galaxies within the innermost bin (Extended Data Figure\,\ref{fig:clustering}, left), resulting in an overdensity of $\delta\approx30$.
Assuming a real space LRD-galaxy two-point correlation function of the form\cite{Eilers2024} $\xi_{\textrm{LG}} = (r/r_{0}^{\textrm{LG}})^{-\gamma}$ with $\gamma=2.0$, we calculate a best-fit cross-correlation length of \rcrosscorr\ (see Methods).
Our result on $r_{0}^{\textrm{LG}}$ is lower but still comparable to the recent quasar clustering measurement\cite{Eilers2024} at $\langle z\rangle=6.25$ with a cross-correlation length of $r_0^{\textrm{QG}}\approx9.1_{-0.6}^{+0.5}\,\textrm{h}^{-1}\,\textrm{cMpc}$. 

Assuming that galaxies and LRDs trace the same underlying dark matter overdensities\cite{GarciaVergara2017}, we adopt the recent estimate of the galaxy auto-correlation length\cite{Eilers2024} of $r_0^{\textrm{GG}}\approx4.1\,\textrm{h}^{-1}\,\textrm{cMpc}$, to estimate the LRD auto-correlation length, $r_0^{LL}$, and the minimum mass of dark matter halos hosting $z\approx7.3$ LRDs:  \logMmin\ (see Methods).  
Our result is broadly consistent with similar minimum halo mass estimates for (luminous) quasars\cite{Arita2023,Eilers2024} $\log_{10}(M_{\textrm{halo,min}}/M_\odot)\approx12.3-12.7$ at $z\gtrsim6$, which would imply that quasars and LRDs are indeed hosted by comparable mass dark matter halos, traced by similar overdensities of galaxies, and thus drawn from the same underlying population.
However, we note that the uncertainties also encompass dark matter halo masses of $\log_{10}(M_{\textrm{halo,min}}/M_\odot)\approx11.5$, akin to recent measurements\cite{Arita2025} of broad-line AGN at $z\sim5.4$.
To constrain the duty cycle of LRDs, the fraction of cosmic time a galaxy spends in an LRD phase, we assume that they temporarily subsample their hosts and so their number density can be expressed as $n_{\textrm{LRD}} \simeq t_{\rm{LRD}}/t_{\rm{H}}(z) n_{\textrm{halo,min}} = f_{\rm{duty}} n_{\textrm{halo,min}}$, where $n_{\textrm{halo,min}}$ is the number density of dark matter halos with $M > M_{\textrm halo, min}$.
Adopting the LRD abundance\cite{Kokorev2024} at $z\approx7.5$ ($\log_{10}(n_{\textrm LRD}/\,\textrm{cMpc}^{-3}) = -5.58\pm0.44$), we calculate a duty cycle of \fD.
%
Taken at face value, the median duty cycle implied by our measurement is unphysical given the constraints of our cosmological model; LRDs vastly outnumber dark matter halos with the median dark matter halo mass. 
This scenario was discussed in a recent paper\cite{Pizzati2025}, leading the authors to conclude that LRDs and comparably luminous quasars cannot only be hosted in dark matter halos with similar masses.
%
Given 1) our small sample size of a single LRD, 2) that we have likely underestimated our error bars by neglecting cosmic variance, and 3) potential systematics associated with assuming a power-law correlation function\cite{Pizzati2024}, we do not believe our results imply a departure from the standard cosmology. 
Instead, we consider it more likely that the majority of LRDs are hosted in less massive halos than our median result suggests. At $z\sim7.3$ dark matter halos with $\log_{10}(M_{\textrm{halo,min}}/M_\odot)\lesssim 11.6$, well within our uncertainties, would already result in physical duty cycles, $f_{\textrm{duty}}\lesssim 100\%$. However, the clustering of \agn\ underlines that LRDs can also be found in more massive dark matter halos.  

%

\begin{figure}[ht!]
\centering
\includegraphics[width=0.8\textwidth]{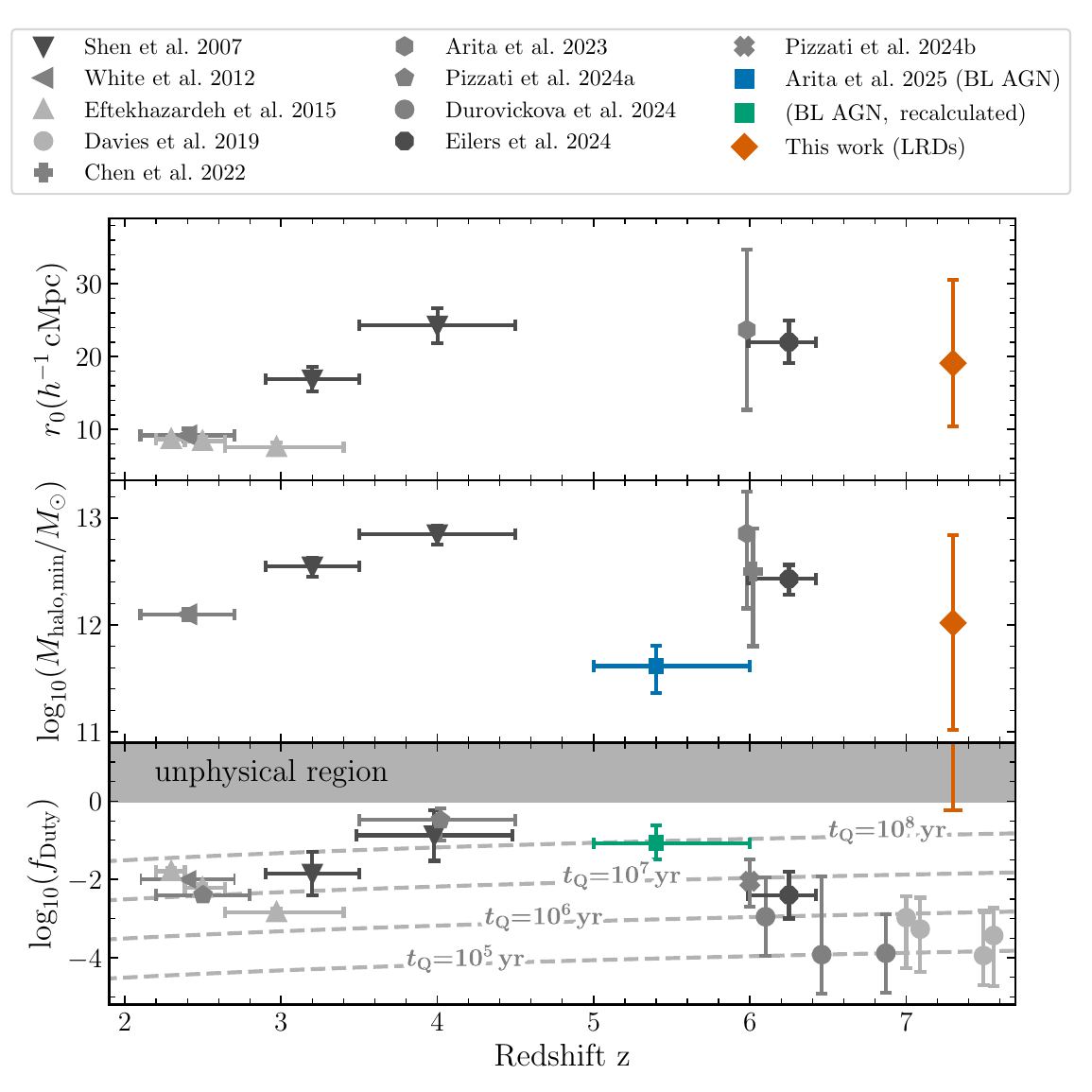}

\caption{\small \textbf{Auto-correlation length, minimum dark matter halo mass, and duty cycle for accreting SMBHs as a function of redshift.}
In the top and middle panel we compare the auto-correlation length and minium dark matter halo mass of our results (orange diamond) on LRDs with  clustering measurements of quasars\cite{ShenYue2007, White2012, Eftekharzadeh2015, Arita2023, Pizzati2024, Eilers2024, Pizzati2024a} and broad-line AGN\cite{Arita2025} (BL AGN).
The bottom panel summarizes literature results on the duty cycle for accreting SMBHs and also includes measurements from quasar proximity zones\cite{ChenHuanqing2022} (gray cross) and Ly$\alpha$ damping wings\cite{Davies2019, Durovcikova2024} (filled circles).
The green square marks our recalculation of the duty cycle estimate for BL AGN\cite{Arita2025} (see Methods).
The duty cycle, $f_{\textrm{duty}}$, should be interpreted as the fraction of cosmic time a galaxy spends in a LRD or quasar phase and is related to the active lifetime via $t_{\textrm{Q}}=f_{\textrm{duty}}\cdot t_{\textrm{H}}(z)$, indicated as gray dashed lines in the bottom panel.
Vertical error bars denote the $1\sigma$ statistical uncertainty on the measurment, whereas horizontal error bars denote the redshift range of the underlying data sample.
}
\label{fig:clustering_overview}
\end{figure}

%
In Figure\,\ref{fig:clustering_overview} we place our clustering results in context with the redshift evolution of auto-correlation lengths, minimum host dark matter halo masses, and duty cycles for UV-luminous quasars and JWST broad-line AGN\cite{Arita2025}. 
Our LRD auto-correlation length and dark matter halo mass measurements are generally consistent with both UV-luminous quasars at $z\approx6$ and JWST broad-line AGN at $z\approx5.5$.  
The high duty cycle suggested by our analysis is in stark contrast with the far lower duty cycle inferred for UV-luminous quasars ($<1\%$ at $z\gtrsim6$) from both quasar-galaxy clustering\cite{Eilers2024, Pizzati2024} and their Ly$\alpha$ forest proximity zones\cite{Davies2019, Durovcikova2024}.  
It is also slightly larger compared to the duty cycle implied by the clustering of galaxies around $z\sim 5.4$ broad-line AGN \cite{Arita2025} (recalculated, see Methods). 

In this work we have introduced the $z=7.3$ LRD \agn. It is one of the most luminous LRDs and if attenuated by dust, its intrinsic properties would be akin to UV-luminous quasars. Our spectroscopic LRD-galaxy clustering analysis places this source in a $\sim10^{12}\,\textrm{M}_\odot$ dark matter halo, indicating that it could be drawn from the same underlying population as UV-luminous quasars.
Generalizing our clustering measurement to the LRD population, we argue that LRDs are populating the massive end of the dark matter halo distribution ($\gtrsim10^{11.5}\,\textrm{M}_\odot$), resulting in high duty cycles of $f_{\textrm{duty}}\sim 100\%$, in contrast to UV-luminous quasars ($f_{\textrm{duty}}\sim 1\%$). 
Hence, we propose that the bulk of massive SMBH growth at high redshift occurs in long-lived phases wherein the SMBHs appear as LRDs, and that they appear as UV-luminous quasars just a short fraction ($f_{\textrm{duty}}\sim 1\%$) of cosmic time.
In this picture, the factor of $\sim 100$ luminous LRD:quasar abundance ratio, naturally results from the fraction of time SMBHs spend in each phase \cite{Davies2019,Satyavolu2023,Eilers2024, Jahnke2025}, which alleviates the tension between the short inferred lifetimes of UV-luminous quasars relative to the time required to grow their  $10^9\,\textrm{M}_\odot$ SMBHs in less than 1\,Gyr after the Big Bang\cite{Inayoshi2020}.
Although highly suggestive, an important caveat is that the UV-luminous quasars we compare with in Figure~\ref{fig:clustering_overview} have $L_{\textrm bol}$ five to 20 times brighter than that of \agn, which is itself uncertain due the reddening correction, and there is likely a dependence of clustering, halo mass, and duty cycle on $L_{\textrm bol}$\cite{Pizzati2025}. Furthermore, our analysis, based on a handful of galaxies in a single LRD field, has large statistical errors, which are likely underestimated owing to cosmic variance and the simplicity of our modeling. Nevertheless, our study presents highly suggestive evidence for strong LRD clustering and high duty cycles, which provides a compelling motivation to further pursue precise LRD clustering measurements, to unravel the nature of this puzzling population and its connection to quasars and SMBH growth.

\newpage

\newpage

\clearpage

\section*{Methods}

\subsection*{Data \& data reduction}
The data presented in this work were taken as part of the JWST program GO 2073 ``Towards Tomographic Mapping of Reionization Epoch Quasar Light-Echoes with JWST''. The goal of the program is to understand the timescales on which $z\gtrsim 7$ quasars grow by tomographically mapping the extended structure of the ionized region around those quasars, their ``light-echos''\cite{Schmidt2019}. For this purpose, one measures the transmission of flux in the spectra of background galaxies, which serve as line-of-sight tracers for the ionized region. 
The program GO 2073 was primarily designed to identify galaxies  at the redshift and in the background of two high redshift quasars. We therefore first collected JWST/NIRCAM photometry in the two quasar fields for galaxy candidate selection and then spectroscopically followed-up the candidates with the NIRSpec/MSA in the same cycle. \agn\ was originally selected as a priority 1 galaxy candidate in the field of quasar J1007+2115 and serendipitously discovered during the spectroscopic follow-up campaign.
For all cosmological calculations we adopt a concordance cosmology\cite{Hinshaw2013} with $H_{0}=70\,\textrm{km}\,\textrm{s}^{-1}\,\textrm{Mpc}^{-1}$, $\Omega_{\Lambda}=0.7$, and $\Omega_{M}=0.3$.

\subsubsection*{Photometric data}
The source \agn\ was discovered in a $\sim5^\prime {\times} 6^\prime$ field centred on the redshift $z\!=\!7.51$ quasar J1007+2115 \cite{YangJinyi2020}.
The JWST/NIRCam observations cover the F090W, F115W, F277W, and F444W filter bands to enable galaxy dropout selection. The mosaic is built from two pointings employing a FULLBOX 6 dither pattern using the MEDIUM 8 readout pattern.  With 3736.4\,s of exposure time per pointing and SW/LW filter pair, the NIRCam/imaging was charged a total science time of 14952\,s.

\noindent We downloaded the data using the \texttt{jwst\_mast\_query}\footnote{\url{https://github.com/spacetelescope/jwst_mast_query}} python package. Our image data reduction was carried out using version 1.6.3 of the JWST Science calibration pipeline\footnote{\url{https://jwst-pipeline.readthedocs.io/en/latest}} (\texttt{CALWEBB}; CRDS context \texttt{jwst\_1046.pmap}).

During the reduction, we executed a range of additional steps.
After \texttt{CALWEBB} stage 1, we performed 1/f-noise subtraction on a row-by-row and column-by-column basis for each amplifier\cite{Schlawin2020}. We continued by running stage 2 of the pipeline on these 1/f-noise subtracted images.
Detector-dependent noise features were apparent in the stage 2 outputs. In order to remove them, we constructed master background images for each filter and detector combination by median filtering all available exposures in our program. Scaled master backgrounds were then subtracted from our stage 2 outputs. 
Proper alignment of individual detector images proved challenging due to the low number of point sources in reference catalogues. Therefore, we devised a multi-stage alignment process. 
We began by running \texttt{CALWEBB} stage 3 on the F444W dither groups with the largest overlap. This resulted in four F444W sub-mosaics corresponding to combined images for each detector and each of our two pointings. 
These four sub-mosaics were then aligned to the positions of known Gaia stars in the field using \texttt{tweakwcs}\footnote{\url{https://github.com/spacetelescope/tweakwcs}}. Choosing one tile as a reference, we iteratively aligned the other three tiles using the Gaia star catalogue and point sources from the reference tile. 
The aligned F444W sub-mosaics were resampled into one mosaic with a pixel scale of $0.03^{\prime\prime}$. As a final step, the full mosaic is then aligned to the full set of Gaia reference stars in the field.  
With the F444W full mosaic as a reference image, we align the \texttt{CALWEBB} stage 2 output files to a common reference frame, (including the F444W images). 
Then we create association files for the aligned images of each filter and run \texttt{CALWEBB} stage 3 (with the \texttt{tweakregstep} disabled) to create mosaics with a pixel scale of $0.03^{\prime\prime}$ for each filter band. The mean alignment accuracy measured from the full mosaics and the Gaia reference star catalogue is $0.024^{\prime\prime}$, $0.018^{\prime\prime}$, $0.018^{\prime\prime}$, $0.011^{\prime\prime}$ for the F090W, F115W, F277W, F444W bands, respectively. 
We estimate that our full mosaics cover a field of view of $\sim29\,\textrm{arcmin}^2$.

In order to calculate source photometry, we begin by resampling the mosaics to a common WCS grid. Then we convolve the resampled F090W, F115W, and F277W mosaics to the lower resolution of the F444W filter. The point-spread function (PSF) convolution kernels are generated empirically from point sources in the field. 
To extract photometry, we use the SExtractor\cite{Bertin1996} software. Source detection is performed on an inverse variance weighted signal-to-noise image stack of the four mosaics. We calculate Kron\cite{Kron1980} aperture photometry on all detected sources using a Kron parameter of 1.2. The fluxes measured in these small apertures are then aperture corrected in two steps, equivalent to common procedures\cite{Bouwens2016}. The correction for the PSF wings uses empirically generated PSFs based on point sources in the field.  

To improve our high-redshift galaxy selection, we obtained deep r-band and i-band dropout images with LBT/LBC\cite{Giallongo2008}. The observations were performed in binocular mode, with individual exposures for all images set to $\sim$180 s to minimize the effects from cosmic rays and the saturation of bright stars in the field. In total, the on-source time is 9.5 hours for both r-band and i-band.
We processed the LBC data using a custom data reduction pipeline named {\tt PyPhot}\footnote{\url{https://github.com/PyPhot/PyPhot}}. The pipeline introduces standard imaging data reduction processes including bias subtraction, flat fielding, and sky background subtraction.  
The master bias and flat frames were constructed using the sigma-clipped median on a series of bias and sky flats, respectively.  For i-band, we further corrected fringing by subtracting off a master fringe frame constructed from our science exposures. The sky background was estimated using {\tt SExtractor}\cite{Bertin1996} after masking out bright objects in the images. In addition, we masked cosmic rays using the Laplacian edge detection algorithm\cite{vanDokkum2001}. Finally, we aligned all individual images to GAIA DR3 and calibrated the zero points with well detected point sources in the Pan-STARRS\cite{Chambers2016} photometric catalogues. 
After the image processing, we created the final mosaics for each band using {\tt SCAMP}\cite{Bertin2006} and {\tt Swarp}\cite{Bertin2002}. The pixel scales of the final mosaics are $0.224 ''/$pixel for the LBC images.

\subsubsection*{Galaxy candidate selection}
Based on the JWST/NIRCam and ground-based photometry we devised a photometric galaxy candidate selection, targeting galaxies in the field of quasar J1007+2115 and in the background. 
The target redshift interval for this quasar field is $7.43 < z < 9.09$. The maximum redshift is chosen so that the foreground quasar is still within the Ly$\beta$ forest of the background galaxy. The minimum redshift is chosen so that galaxies with $-3000\,\textrm{km}\,\textrm{s}^{-1}$ line-of-sight velocity relative to the quasar are included to allow for a quasar-galaxy clustering analysis. 
Our galaxy selection procedure is largely informed by the properties of high redshift galaxies as presented in the JAGUAR catalog\cite{Williams2018}. 
We first require a signal-to-noise ratio (SNR) of $2.0$ for source detections in the F115W and F277W bands. We further clean our photometric sample by removing sources with F115W to F277W colors that are outside the range of $-1.5$ and $1.5\,\textrm{mag}$. This does not affect galaxies in our target redshift range according to the JAGUAR catalog.
We use a probabilistic dropout selection based on the F090W to F115W color as the main selection criterion. 
Based on the F090W to F115W colors of JAGUAR mock galaxies, we assign each color value with a purity. The purity is the fraction of mock galaxies with that color value within the target redshift range compared to all mock galaxies with that color value.
Each galaxy candidate is then assigned a purity value and the full candidate list is ordered in decreasing purity. 
We continue by grouping the first 100 sources in priority class 1 and then proceed with 100, 100, 700, 1000 sources for classes 2, 3, 4, and 5, respectively. All remaining sources are assigned priority class 6.
The broad priority classes are used in our spectroscopic MSA design with the eMPT tool\cite{Bonaventura2023}.
We performed visual inspection of our candidates, utilizing the JWST NIRCam and ground-based LBT photometry. Additionally, photometric redshift were calculated with \texttt{bagpipes}\cite{Carnall2018}\footnote{\url{https://bagpipes.readthedocs.io/en/latest/}} using all photometric information  to guide the process.
As we do not expect significant source flux in the NIRCam F090W filter ($\lambda=0.795-1.005\,\mu\textrm{m}$) for high-redshift sources, galaxy candidates with a significant F090W detection were demoted to a lower priority class. 
This procedure prioritizes sources in the targeted redshift range, but does not result in a hard lower redshift cutoff.
The resulting candidate list was used as input for the spectroscopic follow-up observations.

\subsubsection*{Spectroscopic data}

Galaxy candidates were observed in two NIRSpec/MSA pointings with the PRISM/CLEAR disperser filter combination, providing continuous spectra from $0.6\,\mu{\textrm{m}}$ to $5.3\,\mu{\textrm{m}}$ with a resolving power of $R\sim30 - 330$.
The MSA masks were designed using the eMPT tool\cite{Bonaventura2023}\footnote{\url{https://github.com/esdc-esac-esa-int/eMPT_v1}} with the goal to maximize the coverage of priority 1 sources from our candidate list.
Before designing the mask, we visually inspected all priority 1 through 5 candidates, removing sources where the photometry was obviously affected by image artefacts. Of all priority 1 (2, 3) candidates, 78 (64, 52) passed this step. We were able to assign MSA slits to 52 candidates of the 124 candidates that were covered by the area of the two final MSA pointings. Of all 56 (35, 33) priority 1 (2, 3) targets 34 (9, 9) received slits, resulting in a targeting completeness of 61\%, 26\%, 27\%, respectively.
Each pointing was observed using the standard 3 shutter slitlet nod pattern with 55 groups per integrations and 2 integrations  per exposure and read out using the NRSIRS2RAPID pattern. This resulted in a total exposure time of $4902\,\textrm{s}$ per pointing.

The raw rate files were downloaded from STScI using \texttt{jwst\_mast\_query} and then reduced with a combination of the \texttt{CALWEBB} pipeline and the \texttt{PypeIt}\cite{Prochaska2020}\footnote{\url{https://pypeit.readthedocs.io/en/release/}} python package. 
The spectroscopic reduction was carried out using version 1.13.4 of the JWST Science calibration pipeline\footnote{\url{https://jwst-pipeline.readthedocs.io/en/latest}} (\texttt{CALWEBB}; CRDS context \texttt{jwst\_1188.pmap}. Only \agn has been re-reduced recently with the CRDS context jwst\_1215.pmap.
\texttt{PypeIt} is a semi-automated pipeline for the reduction of astronomical spectroscopic data, which includes JWST/NIRSpec as a supported spectrograph. 
The rate files were first processed with the \texttt{CALWEBB} Spec2Pipeline skipping the \texttt{bkg\_subtract}, \texttt{master\_background\_mos}, \texttt{resample\_spec}, and \texttt{extract\_1d} steps, which were then performed using PypeIt. 
We used difference imaging with PypeIt for background subtraction and then co-added the 2D spectra according to the nod pattern, super-sampled on a finer pixel grid (factor 0.8).
The final 1D spectra for all sources were optimally extracted from their 2D coadded spectra.
To ensure that the physical properties determined for \agn\ are accurate, we additionally apply an absolute flux correction to the \agn\ spectrum. By calculating the synthetic NIRCam F115W, F277W, and F444W band fluxes from the final \agn\ spectrum, we determine an empirical flux correction factor of 1.21. Figure\,\ref{fig:discovery} (bottom panel) shows the resulting excellent agreement of the photometry with the spectrum in the three detection bands.

\subsection*{Analysis}

\subsubsection*{Galaxy discoveries}
The goal of the Cycle\,1 program GO 2073 was to discover bright galaxies at the redshifts of two high-redshift quasars and beyond. 
We followed up galaxy candidates spectroscopically using the NIRSpec MSA with the PRISM disperser.
Here, we only report on the galaxy discoveries related to the environmental analysis of \agn, whereas  the full sample will be presented in a future publication on this Cycle\,1 program.

Among all 52 galaxy candidates that were followed up in the J1007 field, we discovered 8 galaxies within a line-of-sight velocity difference of $\left| \Delta v_{\textrm{LOS}} \right| = 2500\,\textrm{km}\,\textrm{s}^{-1}$ relative to \agn.
These were spectroscopically identified by their \oiiibr\,$\lambda\lambda4960.30,5008.24$ emission line doublet.
It is important to note that these galaxies were not specifically targeted to study the environment of \agn.
Their redshifts have been determined by fitting for the redshift of the \oiiibr\,$\lambda\lambda4960.30,5008.24$ doublet in the galaxy spectra. The spectra were modelled using a power-law continuum component and one Gaussian component for each of the doublet emission lines, whose redshift and FWHM were coupled.    
Extended Data Table\,\ref{tab:source_info} provides the galaxy coordinates, their $z_{\textrm{OIII}}$  redshift, and their NIRCam fluxes.
For each galaxy we further calculate the line-of-sight velocity distance $\Delta v_{\textrm{LOS}}$, the angular separation (both in arcseconds and in proper kpc), 
and their absolute UV magnitudes $M_{\textrm{UV}}$, approximated by the absolute magnitude of the F115W filter band. These properties are summarized in Extended Data Table\,\ref{tab:galaxy_info} and are used in our analysis of the LRD-galaxy cross-correlation measurement.
We note that the faintest galaxy we could identify in our spectroscopic sample has a UV magnitude of $M_{\textrm{UV}}\approx -18.9$, which we adopt as our spectroscopic UV detection limit for galaxies in the following clustering analysis.
%


\subsubsection*{Spectroscopic Analysis of J1007\_AGN}
In many LRDs, rest-frame UV lines (e.g., \siiv, \civ, \ciii, \mgii) are uncharacteristically weak for type-1 AGN, fully absent, or more consistent with photoionization from massive stars\cite{Akins2025}. 
In contrast, the \agn\ spectrum shows evidence for rest-frame UV lines, including a weak detection of \niv\ $\lambda1486$ and \civ  emission ($\textrm{SNR}_{\textrm{peak}}\approx3$) and a possible detection of the \mgii\ line ($\textrm{SNR}_{\textrm{peak}}\approx2$). While \niv\ $\lambda1486$ emission is rarely detected in AGN or quasars\cite{Jiang2008}, recent observations have reported nitrogen-enriched gas in a range of $z\gtrsim6$ galaxies\cite{Cameron2023, Topping2024, Castellano2024} and LRDs\cite{Juodzbalis2024, Tripodi2024}.
Curiously, we also observe a downturn of the continuum flux blueward of \civ\ in combination with the absence of strong \Lya\ emission, which might indicate the presence of a strong broad absorption line system. However, the 
low resolution of our present data in this wavelength range precludes further interpretation.

We use the \textsc{Sculptor}\cite{Schindler2022} python package to model the \agn\ rest-frame optical spectrum in the wavelength range 31000\,\AA\ to  43000\,\AA\ ($3754\,$\AA\ to $5207\,$\AA\ rest-frame) with a combination of a power-law for the continuum and Gaussian profiles for the emission lines. 
The wavelength range is set to not include too much flux redward of the \oiiibr\ lines, where the power-law slope of the continuum is expected to change\cite{VandenBerk2001}.
The \oiiibr\,$\lambda4960.30$, \oiiibr\,$\lambda5008.24$, 
\oiiibr\,$\lambda4364.44$, \heii\,$\lambda4687.02$,  and \neiii\,$\lambda3869.85$ lines are modelled with one Gaussian component each. 
We approximate the \Hb\,$\lambda4862.68$ line and \Hg\,$\lambda4341.68$ lines with two Gaussian components 
each, while the \Hd\,$\lambda4102.89$ line is modeled with a single (narrow) component.
Due to the low signal-to-noise ratio of the \neiii\,$\lambda3968.58$ emission lines, we decided against including it in the model and accordingly mask out its contribution.
 
Modelling the rest-frame optical continuum is complicated by contributions from a multitude of atomic and ionic iron emission lines blending into a pseudo-continuum\cite{Boroson1992, Vestergaard2001}. 
Adding an iron pseudo-continuum to the model produces the same results, because the amplitude is effectively set to zero. 
This indicates that the low-resolution and modest SNR of our spectrum cannot constrain the iron pseudo-continuum at present and hence we do not include it in our fit.
 
We expect the widths and redshifts of some of the narrow or broad emission lines to be correlated. In order to better decompose these line components, we couple the redshift and FWHM of the \Hb, \Hg, and \Hd\ narrow line components to the \oiiibr\,$\lambda4960.30$, \oiiibr\,$\lambda5008.24$, and
\oiiibr\,$\lambda4364.44$ lines.
Additionally, the redshift and FWHM of the \Hb\ and \Hg\ broad line components are also coupled together. 
To constrain the model of the low signal-to-noise detection of the \heii\ line, we couple its redshift to the redshift of the narrow emission lines.

AGNs and type-1 quasars excite both \Hg\ and \oiiibr\,$\lambda4364.44$ emission\cite{VandenBerk2001}, which are usually blended due to the broad nature of the lines.
Unfortunately, the quality of our spectrum does not allow us to uniquely decompose the individual line contributions without coupling the redshifts and widths as described above (Figure\,\ref{fig:discovery}).

We sample the full parameter space of the model fit using \texttt{emcee}\cite{ForemanMackey2013}. As results we quote the median value of the fit parameter and report the 68th percentile range of the fit posterior as our $1\sigma$ uncertainty. 
Our model fits account for the low resolution of the PRISM observations by convolving the model spectrum with the wavelength dependent dispersion curve provided by STScI. 
As a consequence all fitted line widths are intrinsic and do not require a resolution correction.
At the wavelength of the narrow \oiiibr\,$\lambda5008.24$, this dispersion curve predicts a resolution of $R\approx193$ ($\textrm{FWHM}\approx1550\,\textrm{km}\,\textrm{s}^{-1}$). Without accounting for the line spread function we measure a FWHM of $\sim1150\,\textrm{km}\,\textrm{s}^{-1}$ for the \oiiibr\,$\lambda5008.24$ line. Hence, the actual spectral resolution is better than the nominal spectral resolution of the MSA PRISM. This is not surprising as the nominal resolution assumes flat illumination of the MSA slits, whereas our target only partially fills the slit (see Figure\,\ref{fig:discovery}).
For our fit procedure we assume a spectral resolution of $R\approx273$ ($\simeq1100\,\textrm{km}\,\textrm{s}^{-1}$) at \oiiibr\,$\lambda5008.24$ and scale the dispersion curve accordingly. 
We have tested the fit results against varying assumptions on the adopted spectral resolution at \oiiibr\,$\lambda5008.24$ ($R\approx268$ $\simeq1120\,\textrm{km}\,\textrm{s}^{-1}$; $R\approx278$ $\simeq1080\,\textrm{km}\,\textrm{s}^{-1}$).
Across the three adopted resolutions the median fit results all agree well within their 16 to 84 percentile ranges.

Extended Data Table\,\ref{tab:spec_info} summarizes the main source properties of \agn\ calculated from the fit in addition to the source redshift, which is determined from a separate line fit to the \oiiibr\,$\lambda4960.30$ and \oiiibr\,$\lambda5008.24$ lines. 
The absolute magnitude at $1450$\,\AA, $M_{1450}$ is directly calculated from the average spectral flux in a $50$\AA\ window around rest-frame $1450$\AA.
We note that fluxes and line widths of the \Hb\ and \Hg\ emission line components face considerable uncertainties. While our model fit treats coupled narrow and broad emission line widths consistently, the low resolution still results in a large degeneracy between the narrow and broad components. These uncertainties limit the use of the Balmer decrement to constrain \agn's\ dust attenuation and are carried over to the derived properties (e.g., the SMBH mass).
\subsubsection*{Nature of the continuum emission}
The width of the \Hb\ broad-line component ($\textrm{FWHM}_{\Hb,\textrm{broad}}/(\textrm{km}\,\textrm{s}^{-1}) > 3000$ leaves little doubt that our source is a bona-fide type-1 AGN.
However, rest-frame UV emission lines (\civ, \siiv, \mgii, \ciii), which are expected to be strong in type-1 AGN appear weak and the continuum beyond $\sim3500$\AA\ has an unusually red slope (\AlphaOpt; $f_{\lambda} \propto \lambda^{\alpha_{\rm{OPT}}}$).
The resulting red rest-frame optical colour (F277-F444W$\gtrsim1.5$) and the source's compact nature (see Extended Data Figure\,\ref{fig:phot_comp}) mark this source as belonging to the population of LRDs\cite{Matthee2024, Greene2024, Akins2024}. These are compact, red (rest-frame optical) sources discovered in JWST imaging data.
Spectroscopic surveys have identified broad \Ha\ or \Hb\ line components in photometrically selected compact, red sources, suggesting that a significant fraction are (type-1) AGN ($\sim60\%$\cite{Greene2024}). 
However, their lack of strong rest-frame UV emission lines typical for AGN and their unusual V-shaped continuum (blue rest-frame UV and red rest-frame optical) makes it challenging to classify them in the AGN unification paradigm.
One hypothesis is that the continuum emission is the superposition of a significantly dust attenuated AGN continuum and much weaker scattered intrinsic AGN emission, responsible for the blue rest-frame UV slope\cite{Greene2024, Akins2024}, similar to lower-redshift dust reddened quasars\cite{Stepney2024}.
Alternatively, the spectrum could be explained with dust attenuated AGN emission in combination with unattenuated\cite{Brooks2025} stellar light dominating the rest-frame UV emission\cite{Greene2024, Akins2024}.
It has also been proposed\cite{Inayoshi2025} that dense gas is responsible for the Balmer break and absorption features often observed in LRDs\cite{deGraaff2025,Juodzbalis2025, Naidu2025}.

We here consider scenarios that attribute the shape of the LRD continuum emission to some level of dust attenuation between the observer and the broad line region of the AGN. In this scenario, our measurements of the bolometric luminosity, black hole mass and Eddington luminosity ratio derived from the rest-frame optical spectrum will be biased low.
In order to estimate the level of dust attenuation, we first measure the Balmer decrement from the narrow \Hb\ and \Hg\ line (see tabulated values\cite{Osterbrock2006} for $T=10^4\,\textrm{K}$; $n_{e}=10^6\,\textrm{cm}^{-3}$). 
From the flux ratio of these Balmer lines, we estimate a dust attenuation of $A_V = {5.67}_{-6.49}^{+6.07}\,\textrm{mag}$. Different assumptions on the spectral resolution lead to similarly high results ($R\approx268$: $A_V = {4.58}_{-6.63}^{+5.15}\,\textrm{mag}$; $R\approx278$: $A_V = {6.61}_{-5.90}^{+6.73}\,\textrm{mag}$).
%

The large uncertainties are due to the degeneracy with the broad Balmer lines in our fit. 
We note that we do not derive a Balmer decrement from the broad-line components. 
Their flux ratios are known to deviate from the theoretical expectations due to the changing conditions and high densities of the broad-line region \cite{Korista2004}.
Using the \Hd\ line for the Balmer decrement leads to  negative, and hence unphysical, dust attenuation values. 
We have modeled the marginally resolved (3-4 pixel) and detected line (SNR$\lesssim4$) with only one narrow component. 
A non-negligible broad-line component may explain this inconsistency. However, given the already large uncertainties on the components of the other Balmer lines, we do not believe that a two-component fit for \Hd\ will lead to meaningful results.

Alternatively, we model the spectrum of \agn\ in line-free windows with two continuum models.
First, we model the continuum emission with a combination of an attenuated power-law model $f(f_{2500}, \alpha_\lambda, A_v)$ and a scattered-light power law $g(f_{2500}, \alpha_\lambda, f_{\textrm{sc}})$. Both power laws are defined by the same intrinsic flux at $2500$\AA, $f_{2500}$, and the same slope, $\alpha_\lambda$. The former model is then attenuated by $A_V$ using a standard attenuation curve\cite{Calzetti2000} and the latter is multiplied by a scattered light fraction $f_{\textrm{sc}}$.
With these models, we perform an MCMC likelihood fit using \texttt{emcee}\cite{ForemanMackey2013}.
We introduce a Gaussian prior on the power law slope with a mean of -1.5 and a standard deviation of $\sigma=0.3$, because a uniform prior fit always preferred extremely steep ($\alpha_{\lambda}<-3$), physically unmotivated, power law slopes. 
All other parameters receive uniform priors. 
Our results indicate that the underlying continuum emission originates from a steep power law (\fitslope; \fitflux), which has been significantly attenuated (\AVa). We find a scattered light fraction of \fitfsc.
Extended Data Figure\,\ref{fig:av_fit} displays the median model along with the 68-percentile posterior range in orange. 
The model can reasonably well approximate the continuum emission within the flux uncertainties.
However, the rest-frame $\lesssim3000$\AA\ region seems slightly underpredicted, whereas the rest-frame $\sim3000-4000\,\textrm{\AA}$ region is slightly overpredicted by this fit.

We also fit a superposition of a galaxy model with a power law slope to the data. 
The power law model is equivalent to the model described above without the scattered light component. We also impose the same prior on the power law slope.
We generate the galaxy model using \texttt{bagpipes}\cite{Carnall2018} and its default stellar population synthesis models.
We adopt a delayed-$\tau$ star formation history, with a uniform prior on the galaxy age ($10$-$735\,\textrm{Myr}$) and $\tau$ ($0.1$-$4.0\,\textrm{Gyr}$).
Stellar masses are allowed within the range of $10^8$ to $10^{12}\,\textrm{M}_\odot$ and the galaxy model receives its own dust attenuation parameter   uniformly sampled within $A_{V,gal}=0.01$-$6\,\textrm{mag}$. 
The same dust attenuation law\cite{Calzetti2000} is applied consistently for both components.
First tests of the galaxy + power law model showed that the stellar metallicity cannot be constrained by the fit, and so we fix it to a value of $0.5$ consistent with the gas-phase metallicity of high-redshift, massive galaxies ($M\approx10^{10}\,\textrm{M}_\odot$).
The blue solid line in Extended Data Figure\,\ref{fig:av_fit} shows the median fit of this model to the spectrum with separate contributions from the galaxy and the power law component highlighted with different line styles. 
The galaxy component dominates the continuum in the rest-frame optical with a clear Balmer break. 
The nominal results for the stellar mass, stellar age and the $\tau$-parameter are $\log(M_*/M_\odot) = {9.83}_{-0.22}^{+0.50}$, $t = {0.45}_{-0.19}^{+0.17}\,\textrm{Gyr}$, and $\tau = {2.22}_{-1.30}^{+1.21}\,\textrm{Gyr}$, respectively. We note that the posteriors show strong degeneracies between the stellar age and the $\tau$-parameter, and that both parameters effectively remain unconstrained. The galaxy component is attenuated by \AVgal.
The AGN, modelled as a power law, is similar steep as in the scattered light model (\fitslopeb) with an amplitude of \fitfluxb. The dust attenuation for the AGN is largely unconstrained, with a nominal median value of \AVb.

Scaling relations between the \Hb\ broad line flux and $5100\AA$ continuum luminosity\cite{Greene2005} imply similar levels of dust attenuation for the AGN continuum and broad line components. 
The line fit to the \agn\ spectrum (Fig.\,\ref{fig:agn_analysis}) results in a value of $L_{\textrm{H}\beta} = 3.89\times10^{42}\,\textrm{erg}\,\textrm{s}^{-1}$, which would correspond to an accretion disk continuum flux of $L_{5100} \approx 4.76\times10^{40}\,\textrm{erg}\,\textrm{s}^{-1}\,\textrm{\AA}^{-1}$, according to the scaling relation. We measure a value of $L_{5100} \approx 2.69\times10^{40}\,\textrm{erg}\,\textrm{s}^{-1}\,\textrm{\AA}^{-1}$ for the continuum component in our spectral fit, close to the value estimated from the scaling relation. 
In the case of the continuum fits, the scattered light model finds a similar continuum luminosity ($L_{5100} \approx 1.69\times10^{40}\,\textrm{erg}\,\textrm{s}^{-1}\,\textrm{\AA}^{-1}$), whereas the model with the galaxy component predicts a much lower luminosity of $L_{5100} \approx 5.35\times10^{39}\,\textrm{erg}\,\textrm{s}^{-1}\,\textrm{\AA}^{-1}$), about a factor of nine lower than inferred from the \Hb\ line luminosity. 
It is a remarkable fact that the \agn\ spectrum comes close to obey the established AGN scaling relations between line and continuum luminosity.
Any contribution from a stellar component to the continuum would decrease the AGN continuum luminosity, essentially breaking with these scaling relations. 
In light of this, we prefer the interpretation of the scattered light model and adopt its posterior attenuation value (\shortAVa) for further analysis, a conservative choice in comparison with the results from the Balmer decrement.

\subsubsection*{Derivation of the SMBH mass and the Eddington ratio}
The AGN nature of LRDs and the association of the observed broad Balmer lines originating in a typical broad-line region (BLR) is highly debated in the literature\cite{Baggen2024, Kokubo2024, Juodzbalis2024, Naidu2025, Rusakov2025, Juodzbalis2025}.
In order to estimate the SMBH mass and Eddington rate of \agn\ we proceed by assuming that the broad lines observed in our spectrum originate from BLR gas and that the single-epoch virial estimators calibrated in the local universe do apply to our source.
Assuming that the BLR emitting gas is in virial motion around the SMBH, we use the line-of-sight velocity width, as measured by the $\textrm{FWHM}$ of the line, to trace the gravitational potential of the SMBH mass $M_{\textrm{BH}}$:
\begin{equation}
M_{\textrm{BH}} \approx f \frac{R \cdot \textrm{FWHM}}{G}\ ,
\end{equation}
where $R$ is the average radius of the line-emitting region and $f$ encapsulates our ignorance on the detailed gas structure, its orientation towards the line of sight and more complex BLR kinematics. 
Correlations connecting the radius $R$ to the continuum luminosity of broad-line AGN\cite{Kaspi2000, Bentz2006} then allow us to rewrite the expression above in terms of direct observables\cite{Vestergaard2002}, i.e. a virial SMBH mass estimator.
Traditionally, these relations estimate the SMBH mass from the FWHM of a line (e.g., the \Hb\ line) and a measure of the continuum luminosity of the source (e.g., the luminosity at $5100$\,\AA, $L_{5100}$).
For quasars that dominate the emission at $5100$\,\AA\ this choice is appropriate. However, it is unclear to what extent the emission in the rest-frame optical is dominated by the AGN or by galaxy light. 
Therefore, we employ a single-epoch virial estimator\cite{Greene2005} that uses the total \Hb\ line luminosity as a proxy for the continuum luminosity for our fiducial SMBH estimate, $M_{\textrm{BH, GH05, LH}\beta}$. 
Additionally, we adopt three different single-epoch virial estimators\cite{Greene2005, Vestergaard2006, Park2012} that use the \Hb\ FWHM and continuum luminosity $L_{5100}$ for comparison. We use these to gauge the systematic uncertainty inherent in this form of SMBH mass measurement. 

To estimate the bolometric luminosity, $L_{\textrm{bol}}$ we apply a typical bolometric correction factor\cite{ShenYue2011} ($L_{\textrm{bol}}= 9.26\cdot L_{5100}$) to estimate the bolometric luminosity from the continuum luminosity at $5100$\,\AA, $L_{5100}$. 
In order to produce consistent results for our fiducial SMBH mass estimator, we  alternatively use the empirical line-to-continuum luminosity relations\cite{Greene2005} to calculate the approximate $L_{5100}$ from $L_{\textrm{H}\beta}$ and then convert $L_{5100}$ to a bolometric luminosity (denoted as $L_{\textrm{bol, H}\beta}$).
We use the appropriate bolometric luminosities to calculate the Eddington luminosity ratios, $\lambda_{\textrm{Edd}} = L_{\textrm{bol}} / (1.26\times10^{38}\,\textrm{erg}\,\textrm{s}^{-1}\,{\textrm{M}_\odot}^{-1} \cdot M_{\textrm{BH}})$, for the different SMBH mass estimates. 
All measured spectral properties along with the luminosities, SMBH masses and  Eddington luminosity ratios are presented in Extended Data Table\,\ref{tab:spec_info}.
Based on the observed spectrum, we find \agn\ to host a SMBH with a mass of \BHM\ with an Eddington luminosity ratio of \lEdd. 
At face value \agn\ hosts a rapidly ($\lambda_{\textrm{Edd}}>0.1$) accreting, relatively massive SMBH ($M_{\textrm{BH}}\approx10^8\,\textrm{M}_\odot$), akin to the least luminous quasars\cite{Fan2023} identified at $z>5.9$ (see Figure\,\ref{fig:agn_analysis}, right panel for a comparison).
The SMBH mass estimates based on different single-epoch scaling relation vary within a factor of 2, in agreement with the expected systematic uncertainties\cite{Vestergaard2006} of $\pm0.43\,\textrm{dex}$ depending on the adopted estimator and model assumptions\cite{Bertemes2025}.

Following our discussion on the nature of the continuum emission, we concluded that the \agn\ spectrum likely suffers from dust attenuation. To estimate the attenuation corrected source properties we adopt the attenuation value of \AVa\ derived from the scattered light model continuum fit.

We correct the model fit realizations to the optical \agn\ spectrum (Fig.\,\ref{fig:agn_analysis})  by applying an attenuation correction for an $A_V$ value that is randomly sampled from the posterior of the scattered light model fit. In this way we consistently propagate both the uncertainties on the line fit and the dust attenuation.

We measure the properties of the dust corrected model fits and recalculate the luminosities, SMBH mass and the Eddington luminosity ratio for our fiducial choice of single-epoch virial estimator\cite{Greene2005} based on the \Hb\ line luminosity.
These results are summarized in Extended Data Table\,\ref{tab:dereddened_prop}.
Accounting for the attenuation, \agn\ reaches quasar-like bolometric luminosities \Lbold and an SMBH mass of \BHMd, now fully overlapping with the quasar distribution\cite{Fan2023} at $z>5.9$ as shown in the right panel of Figure\,\ref{fig:agn_analysis}.

\subsubsection*{Number density estimate}
To place our serendipitous discovery of \agn\ in context with the population of faint high-redshift AGN discovered with JWST, we calculate its approximate number density.
\agn\ was discovered as a priority\,1 galaxy candidate during our spectroscopic follow-up campaign. While the target redshift range for galaxy candidates was approximately $z\approx7.4-9.1$, our permissive selection also selects dropout sources at lower redshifts.
In order to calculate the surveyed volume of our observations we use an inclusive redshift interval of $z=7.2-9.1$. Given our deep NIRSpec PRISM observations, we are confident that we would have detected any similarly bright AGN up to $z=9.1$. The discovery of the \agn\ and many $z\approx 7.2$ galaxies motivates the extension of the lower redshift limit below the nominally targeted redshift range for galaxies. The total survey area we adopt is $16.73\,\textrm{arcmin}^{2}$, the overlap of our NIRCam photometry and NIRSpec MSA spectroscopy.
We derive a survey volume of $V=61402\,\textrm{Mpc}^{3}$. This leads to an approximate source density of $n=1/V\approx1.63\times10^{4}\,\textrm{Gpc}^{-3}$, assuming a total selection completeness of 100\%. However, we already know that our targeting selection completeness for priority\,1 sources is only 61\%. Correcting for this effect, we calculate $n_{\textrm{corr}}\approx1.64/V=2.67\times10^{4}\,\textrm{Gpc}^{-3}$.
In order to compare these number estimates in the context of other samples of faint high-redshift AGN, we calculate luminosity function estimates based on this one source. This is solely for illustration and we do caution that calculating a statistical property from a single source has significant systematic biases due to the small sample size and cosmic variance. 
First, we place \agn\ in the context of the high-redshift UV luminosity function. As a proxy for the absolute UV magnitude, we use the absolute magnitude at $1450$\,\AA\ as measured from the spectrum, $M_{1450}=-19.29$ (see Extended Data Table\,\ref{tab:spec_info}) and choose a bin size of $\Delta M_{1450}=1$. 
With these assumptions, our corrected luminosity function measurement is $\Phi=2.67_{-2.21}^{+6.14}\times10^{4}\,\textrm{Gpc}^{-3}\,\textrm{mag}^{-1}$, where the uncertainties encompass the confidence interval for a Poisson distribution that corresponds to $1\sigma$ in Gaussian statistics.
We compare our estimate with the UV luminosity functions of faint AGN\cite{Harikane2023, Greene2024}, galaxies, and quasars in Figure\,\ref{fig:agn_lf} (left). The right panel of Figure\,\ref{fig:agn_lf} shows our luminosity function estimate converted to bolometric luminosity.
In correspondence with the  literature\cite{Greene2024, Kokorev2024}, we use the bolometric luminosity estimate derived from the \Hb\ line luminosity, $L_{\textrm{bol, H}\beta}$.
These panels show that our luminosity function estimate agrees well with other measurements for faint high-redshift AGN, indicating that the identification of this source in our surveyed volume was likely to be expected. We note that the bolometric luminosity function of these AGN remains orders of magnitude above the best constraints on the bolometric quasar luminosity function \cite{ShenXuejian2020}.

\subsubsection*{LRD-galaxy cross correlation measurement}
The LRD-galaxy cross-correlation function $\chi$ averaged over an effective volume $V_{\textrm{eff}}$ can be related to the LRD-galaxy two-point correlation function $\xi_{\textrm{LG}}$ via
\begin{equation}
    \chi(R_{\textrm{min}}, R_{\textrm{max}}) = \frac{\int \xi_{LG}(R, Z) dV_{\textrm{eff}}}{V_{\textrm{eff}}} \ ,
\end{equation}
equivalent to luminous quasars\cite{Hennawi2006, GarciaVergara2017}.
In this case we have chosen a cylindrical geometry
with radial coordinate $R$, being the transverse comoving distance, and the cylinder height $Z$, the radial comoving distance, 
\begin{equation}
Z = \frac{c}{H(z)}\delta z \ ,
\end{equation}
where $H(z)$ is the Hubble constant at redshift $z$. The volume averaged cross-correlation $\chi(R_{\textrm{min}}, R_{\textrm{max}})$ is calculated in radial bins with bin edges $R_{\textrm{min}}, R_{\textrm{max}}$.
We effectively calculate $\chi(R_{\textrm{min}}, R_{\textrm{max}})$ via 
\begin{equation}
    \chi(R_{\textrm{min}}, R_{\textrm{max}}) = \frac{\langle LG\rangle}{\langle LR \rangle} -1 \ ,
\end{equation}
where $\langle LG\rangle$ is the number of LRD-galaxy pairs in the enclosed cylindrical volume and $\langle LR \rangle$ is the number of random LRD-galaxy pairs in average regions of the Universe.
We consider only \agn\ for the LRD-galaxy clustering measurement here and thus $\langle LG\rangle$ is simply the number of associated galaxies in the volume. 
The random number of galaxies can be expressed in terms of the background volume density of galaxies $\rho_{\textrm{gal}}$ at redshift $z$ in the cylindrical volume $V_{\textrm{eff}}$: 
$\langle LR \rangle = \rho_{\textrm{gal}} V_{\textrm{eff}}$.
Our survey volume is not large enough to allow us to empirically determine the background volume density of galaxies. Hence, we calculate an estimate of $\rho_{\textrm{gal}}$ from the galaxy luminosity function\cite{Bouwens2022}. We integrate the luminosity over a magnitude range of $-30.0 < M_{\textrm{UV}}\le -18.9$, where the faint-end limit corresponds to the faintest spectroscopically identified galaxy in our sample. The resulting galaxy background density is $\rho_{\textrm{gal}}=7.12\times10^5\,\textrm{Gpc}^{-3}$. 
The effective volume in cylindrical geometry can be expressed as 
\begin{equation}
    V_{\textrm{eff}} = \int_{Z_{\textrm{min}}}^{Z_{\textrm{max}}} \int_{R_{\textrm{min}}}^{R_{\textrm{max}}} S(R, Z) 2\pi R {\textrm{d}}R{\textrm{d}}Z \ ,
\end{equation}
where $S(R, Z)$ is the galaxy selection function in terms of both $R$ and $Z$. We decompose $S(R, Z)$ into three  components:
\begin{equation}
S(R, Z) = S_Z(Z) S_R(R) S_T(R) \ ,
\end{equation}
where $S_Z(Z)$ is the redshift dependent completeness, $S_R(R)$ radially dependent coverage completeness, and $S_T(R)$ is the radially dependent targeting completeness. The coverage completeness $S_R(R)$ accounts for the area in the radial annulus that is not covered by our NIRCam observations and NIRSpec/MSA follow-up, whereas the targeting completeness $S_T(R)$ accounts for the fact that only a subset of galaxy candidates in the covered MSA footprints can be followed up with our MSA observations.

Due to the limited number of companion galaxies, we choose three radial bins with bin edges $0.1, 0.6, 2.7,$ and $7.6\,\textrm{h}^{-1}\,\textrm{cMpc}$ ($0.14, 0.86, 3.71,$ and $10.86\,\textrm{cMpc}$). For these bins we calculate a radial coverage completeness of $S_R(R)=1.0, 0.52,$ and $0.26$, respectively, based on the fractional area covered by our NIRSpec observations.   
We limit our selection of galaxies for the clustering measurement by restricting the relative line-of-sight velocity to $\left| \Delta v_{\textrm{LOS}}\right|\le1500\,\textrm{km}\,\textrm{s}^{-1}$ and $\left|\Delta v_{\textrm{LOS}}\right|\le2500\,\textrm{km}\,\textrm{s}^{-1}$, selecting six or alternatively all of the eight galaxies presented in Extended Data Table\,\ref{tab:galaxy_info}. These line-of-sight velocity differences correspond to redshift intervals of $z=7.22-7.30$ or $z=7.20-7.32$, respectively. For simplicity, we conservatively set the redshift dependent completeness $S_Z(Z)$ to a constant $100\%$ over these narrow redshift intervals, providing us with a lower limit on the cross-correlation measurement. 

In addition, our assignment of MSA slits with the eMPT tool introduces a targeting selection function $S_T(R)$ that depends on the candidate's priority class. 
For each priority and radial bin we calculate our targeting completeness $C_{p,R}$ as the fraction of targeted to photometrically selected galaxy candidates in the MSA area. 
All our identified galaxies belong to the priority classes $p=1$ and $2$ (see Extended Data Table\,\ref{tab:galaxy_info}). The relevant completeness values for our three radial bins $R=[1,2,3]$ in increasing distance to \agn \ are $C_{1,1}=1.0$, $C_{1,2}=0.75$, $C_{1,3}=0.57$, and $C_{2,3}=0.22$. 
Based on these values, we can calculate the ``corrected'' number of LRD-galaxy pairs $\langle LG\rangle_{\textrm{r, corr}}$ per radial bin as 
\begin{equation}
  \langle LG\rangle_{\textrm{R, corr}} = \sum_p \frac{\langle LG\rangle_{p,R}}{C_{p,R}} \ ,
\end{equation}
where $\langle LG\rangle_{p,r}$ is the number of LRD-galaxy pairs per radial bin $R$ and priority $p$.
Finally, the targeting selection function is approximated by the fraction of observed to corrected LRD-galaxy pairs
\begin{equation}
S_T(R) = \frac{\langle LG\rangle_{\textrm{R}}}{\langle LG\rangle_{\textrm{R, corr}}} \ ,
\end{equation}
with values of $1$, $0.75$, and $0.32$ for the three radial bins in order of increasing distance.
We present our binned clustering measurement in Extended Data Table\,\ref{tab:cross_corr} for the two samples with relative line-of-sight velocities $\left|\Delta v_{\textrm{LOS}}\right|\le1500\,\textrm{km}\,\textrm{s}^{-1}$ and $\left|\Delta v_{\textrm{LOS}}\right|\le2500\,\textrm{km}\,\textrm{s}^{-1}$, finding an overdensity in the innermost radial bin with $\delta=\langle LG\rangle/\langle LR\rangle -1 \approx 45$ or $\approx26$, respectively.

We now aim to constrain the real-space LRD-galaxy two-point correlation function $\xi_{\textrm{LG}}$ by parameterizing its shape as 
\begin{equation}
\xi_{\textrm{LG}} = \left(r/r_0^{\textrm{LG}}\right)^{-\gamma_{\textrm{LG}}} \ , 
\end{equation}
where $r=\sqrt{R^2+Z^2}$ is the radial coordinate, $r_0^{\textrm{LG}}$ is the cross-correlation length, and $\gamma_{\textrm{LG}}$ is its power-law slope. This form is governed by two parameters. However, our limited statistics do not allow to uniquely constrain both $r_0^{\textrm{LG}}$ and $\gamma_{\textrm{LG}}$. Hence, we assume a value of $\gamma_{\textrm{LG}}=2.0$ for our analysis, which is chosen to allow for comparison with the quasar literature\cite{Eilers2024}. 
We perform a fit to the binned data and sample a Poisson likelihood on a grid. We present the median of the posterior as our results on the cross-correlation length $r_0^{\textrm{LG}}$ in Extended Data Table\,\ref{tab:cross_corr}. Uncertainties reflect the confidence interval for a Poisson distribution that corresponds to $1\sigma$ in Gaussian statistics\cite{Gehrels1986}. 
Figure\,\ref{fig:clustering} (left) presents our binned LRD-galaxy cross correlation measurement with $\left|\Delta v_{\textrm{LOS}}\right|\le1500\,\textrm{km}\,\textrm{s}^{-1}$, including the targeting completeness correction.
The corresponding best-fit cross-correlation length is \rcrosscorr.
Increasing the line-of-sight distance to $\left|\Delta v_{\textrm{LOS}}\right|\le2500\,\textrm{km}\,\textrm{s}^{-1}$ to encompass all discovered nearby galaxies produces a consistent result (see Extended Data Table\,\ref{tab:cross_corr}).

%
Assuming that galaxies and LRDs trace the same underlying overdensities\cite{GarciaVergara2017}, the LRD-LRD auto-correlation $\xi_{\textrm{LL}}$ can be expressed by the galaxy-galaxy auto-correlation $\xi_{\textrm{GG}}$ and the LRD-galaxy cross-correlation $\xi_{\textrm{LG}}$ according to 
\begin{equation}
\xi_{\textrm{LL}} = \xi_{\textrm{LG}}^2/\xi_{\textrm{GG}}. 
\end{equation}
For our analysis we fix the slopes of the auto-correlation functions similar to previous work on quasars\cite{ShenYue2007, Eftekharzadeh2015, GarciaVergara2017, Eilers2024}, assuming $\gamma_{\textrm{GG}}=2.0$ and $\gamma_{\textrm{LL}}=2.0$.
As all of our identified galaxies show prominent \oiiibr\ lines (Extended Data Figure\,\ref{fig:galaxy_discovery}), we adopt a recent measurement of the galaxy auto-correlation length, $r_0^{\textrm{GG}}=4.1\,\textrm{h}^{-1}\,\textrm{cMpc}$, at $\langle z\rangle=6.25$ based on \oiiibr-emitters\cite{Eilers2024}.
Using our cross-correlation length estimate for $\left|\Delta v_{\textrm{LOS}}\right|\le1500\,\textrm{km}\,\textrm{s}^{-1}$, \rcrosscorr, we derive an auto-correlation length of \rautocorr.
We use predictions of a halo model framework (\texttt{halomod})\cite{Murray2013, Murray2021}, 
to link the auto-correlation length to a minimum dark matter halo mass for LRDs at $z\approx 7.3$. Assuming that all LRDs live in dark matter halos with a minimum mass threshold $M_{\textrm{halo, min}}$, we predict the \textcolor{red}{quasar} auto-correlation function using the halo model for different $M_{\textrm{halo, min}}$. For each quasar auto-correlation function, we tabulate the auto-correlation length $r_0^{\textrm{LL}}$ and the cumulative abundance of halos $n_{\textrm{halo,min}}$ with $M>M_{\textrm{halo, min}}$. This allows us to link our estimate $r_0^{\textrm{LL}}$ to a minimum halo mass. 
We calculate a minimum halo mass of \logMmin with a corresponding abundance of \lognhalo. If we compare this value with the $z\approx7.5$ number density of LRDs\cite{Kokorev2024} ($M_{\textrm{UV}}\!=\!-20$, $6.5\!<\!z\!<\!8.5$), $\log_{10}(n_{\textrm{LRD}}/\textrm{cGpc}^{-3})=3.42\pm0.44$, it becomes evident that at a large range of the estimated minimum halo mass would be inconsistent with our current cosmology (see Figure\,\ref{fig:clustering}, right).
Assuming that LRDs subsample their host distribution, one can relate their number density $n_{\textrm{LRD}}$ and the minimum mass host halo abundance $n_{\textrm{halo,min}}$ to their lifetime $t_{\textrm{LRD}}$ using the same arguments as for UV-luminous quasars\cite{Haiman2001, Martini2001},
\begin{equation}
n_{\textrm{LRD}} \simeq \frac{t_{\textrm{LRD}}}{t_{\textrm{H}}(z)} n_{\textrm{halo,min}} \ .
\end{equation}
Here, $t_{\textrm{H}}(z)$ denotes the Hubble time at redshift $z$. We emphasize that the LRD lifetime $t_{\textrm{LRD}}$ is the average time period, in which we observe the source as an LRD.
In this context, the ratio of LRD lifetime to Hubble time is referred to as the duty cycle $f_{\textrm{duty}}=t_{\textrm{LRD}}/t_{\textrm{H}}(z)= n_{\textrm{LRD}}/n_{\textrm{halo,min}}$.
Based on the abundance of LRDs and our inferred cumulative halo abundance $n_{\textrm{halo,min}}$, we calculate a duty cycle of \fD, resulting in a lifetime for $z\approx7.3$ LRDs of \tQ. We accounted for the uncertainties by calculating realizations drawn from our best-fit posterior for $n_{\textrm{halo,min}}$, assuming a log-normal distribution for $n_{\textrm{LRD}}$ and reporting the 16th to 84th percentiles as uncertainties.
We display the redshift evolution of the auto-correlation length, the minimum halo mass and the duty cycle for quasars, broad-line AGN, and our result for LRDs in Figure\,\ref{fig:clustering_overview}.
The unphysical regime for duty cycles is marked in the gray region in the bottom panel. Furthermore, we recalculated the duty cycle for broad-line AGN\cite{Arita2025} (green square, $\log_{10}(f_{\textrm{duty}})=-1.07_{-0.41}^{+0.45}$). In the original work, the authors integrate over a large range of halo mass abundances ($M_{\textrm{halo}}=10^{11}-10^{12}\,\textrm{M}_\odot$) well beyond their derived characteristic halo mass of $\log_{10}(M_{\textrm{halo}}/h^{-1}\textrm{M}_\odot)=11.46$. Adopting their result as a median mass, we calculate a corresponding minimum halo mass of $\log_{10}(M_{\textrm{halo}}/h^{-1}\textrm{M}_\odot)=11.31$. Using the same approach as above, we calculate the halo abundance using \texttt{halomod} at $z=5.4$.
%
%
The bottom panel of Figure\,\ref{fig:clustering_overview} then compares the duty cycles of different sources and from different tracers across cosmic time. At $z>5$ luminous ($M_{1450}\lesssim-26.5\,\textrm{mag}$) quasars have usually low duty-cycles of  $f_{\textrm{duty}}\sim0.1\%$ measured from quasar clustering\cite{Arita2023, Eilers2024, Pizzati2024}, quasar proximity zones\cite{ChenHuanqing2022}, and Ly$\alpha$ damping wings\cite{Davies2019, Durovcikova2024}. 
This is in contrast to $z\sim4$ quasars \cite{ShenYue2007, Pizzati2024a} and to $z\sim5$ broad-line AGN\cite{Arita2025} with $f_{\textrm{duty}}\gtrsim10\%$.
Our analysis results in a duty cycle of \fD. While the median value lies in the unphysical region ($f_{\textrm{duty}}>1$), our lower 16 percentile tail extends well into the region of physical  duty cycles $f_{\textrm{duty}}=0.3-1$. This result is driven by the high number densities of LRDs\cite{Kokorev2024} compared to the number densities of available host dark matter halos in the mass range we infer. 
%

\medskip
\medskip
\medskip

\noindent \textbf{Data availability}\\ 
The JWST data used in this publication from program GO 2073 are publicly available through the MAST portal (\url{https://mast.stsci.edu/portal/Mashup/Clients/Mast/Portal.html}).
All further datasets generated and analysed during this study are available from the corresponding author upon reasonable request. 

\medskip
\noindent \textbf{Code availability}\\ 
The JWST data were in part processed with the JWST calibration pipeline (\url{https://jwst-pipeline.readthedocs.io}). We used to \texttt{tweakwcs} (\url{https://github.com/spacetelescope/tweakwcs}) align the JWST images into mosaics. Ground-based LBT/LBC image reduction was carried out with \texttt{PyPhot} (\url{https://github.com/PyPhot/PyPhot}).

\medskip
\noindent \textbf{Acknowledgements}\\
This work has been supported by the Deutsche Forschungsgemeinschaft (DFG, German Research Foundation) - Project numbers 518006966 (J.-T.S.) and 506672582 (S.E.I.B.), NSF grants AST-2308258 (X.F.) and AST-2513040 (F.W.), and the Cosmic Dawn Center (DAWN), which is funded by the Danish National Research Foundation under grant DNRF140 (K.K.; DAWN Fellowship).
%
%
This paper includes data from the LBT. The LBT is an international collaboration among institutions in the United States, Italy, and Germany. The LBT Corporation partners are: The University of Arizona on behalf of the Arizona university system; Istituto Nazionale di Astrofisica, Italy; LBT Beteiligungsgesellschaft, Germany, representing the Max Planck Society, the Astrophysical Institute Potsdam, and Heidelberg University; The Ohio State University; The Research Corporation, on behalf of The University of Notre Dame, University of Minnesota and University of Virginia.
This work is based [in part] on observations made with the NASA/ESA/CSA James Webb Space Telescope. The data were obtained from the Mikulski Archive for Space Telescopes at the Space Telescope Science Institute, which is operated by the Association of Universities for Research in Astronomy, Inc., under NASA contract NAS 5-03127 for JWST. These observations are associated with program GO 2073.
Support for program GO 2073 was provided by NASA through a grant from the Space Telescope Science Institute, which is operated by the Association of Universities for Research in Astronomy, Inc., under NASA contract NAS 5-03127.

\medskip
\noindent \textbf{Author Contributions}\\
As the first author, J.-T.S. led the general observing program, the data reduction, the presented analysis, and the manuscript preparation.
J.F.H. and F.B.D. led the initial design of the observation program and the preparation for the photometric observations. 
The photometric data reduction and subsequent spectroscopic follow-up observations preparation were led by J.-T.S. and supported by the team (J.F.H., F.B.D., S.E.I.B., F.W., J.Y., R.E., M.M., and R.N.).
Spectroscopic observations were planned by J.-T.S., J.F.H., M.M. and R.N. and reduced by J.-T.S with support from J.F.H and F.B.D.
Visual inspection of the spectroscopic data was carried out by J.-T.S., S.E.I.B, and F.B.D.\
J.-T.S., J.F.H., F.B.D., S.E.I.B., A.B., K.K., A.-C.E., X.F., and E.P. contributed to discussion and the interpretation of the results.
J.-T.S. led the preparation and revision of the manuscript with contributions from all authors at the revision stage.
\medskip
\noindent \textbf{Competing interests} \\
The authors declare no competing interests.


\clearpage

\renewcommand{\figurename}{Extended Data Figure}

\setcounter{figure}{0}

\begin{figure}[ht!]
\centering
\includegraphics[width=0.95\linewidth]{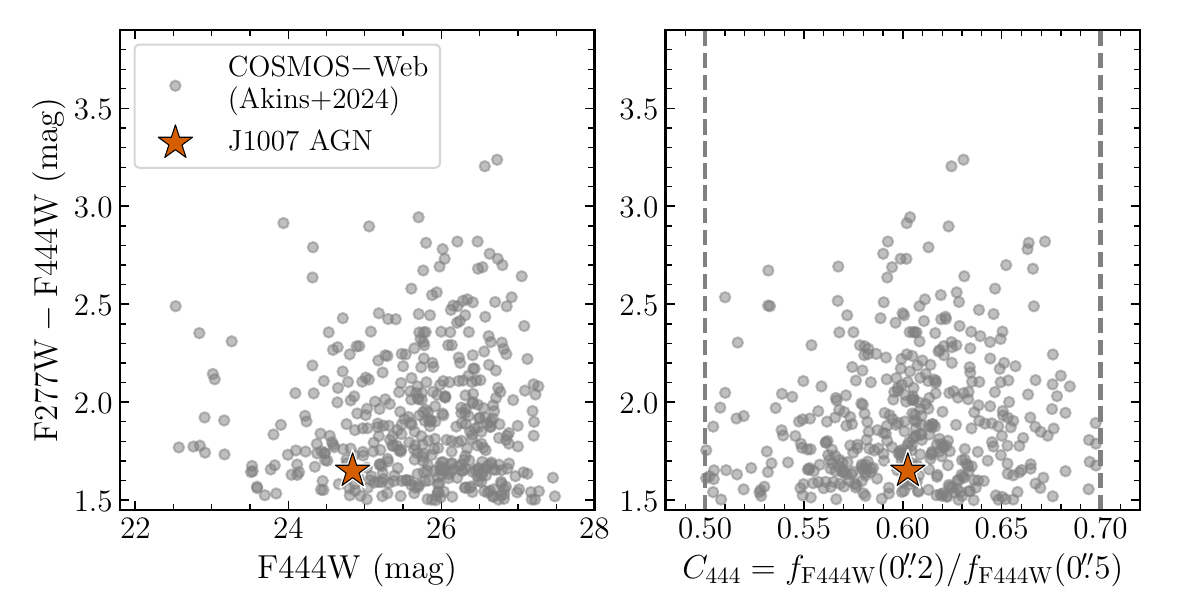}
\caption{\small 
\textbf{Comparison of \agn (orange star) to photometrically selected LRDs from COSMOS-Web\cite{Akins2024} (gray dots).}
\textit{Left:} F277W-F444W colour as a function of F444W magnitude. \agn\ sits above the nominal threshold of F277W-F444W=1.5 for selecting LRDs in the COSMOS-Web sample, while being brighter in F444W than the majority of their photometric LRDs.
\textit{Right:} F277-F444W colours as a function of the F444W compactness $C_{444}$. The compactness is defined as the ratio of the aperture fluxes with  diameters of $0\farcs2$ and $0\farcs5$. \agn\ falls in the centre region, as expected for a source with an unresolved point spread function. 
The dashed grey lines refer to the compactness selection critera ($C_{444}>0.5$ and $C_{444}<0.7$) applied to photometrically selected LRDs\cite{Akins2024}.
}
\label{fig:phot_comp}
\end{figure}

\begin{figure}[ht!]
\centering
\includegraphics[width=0.95\linewidth]{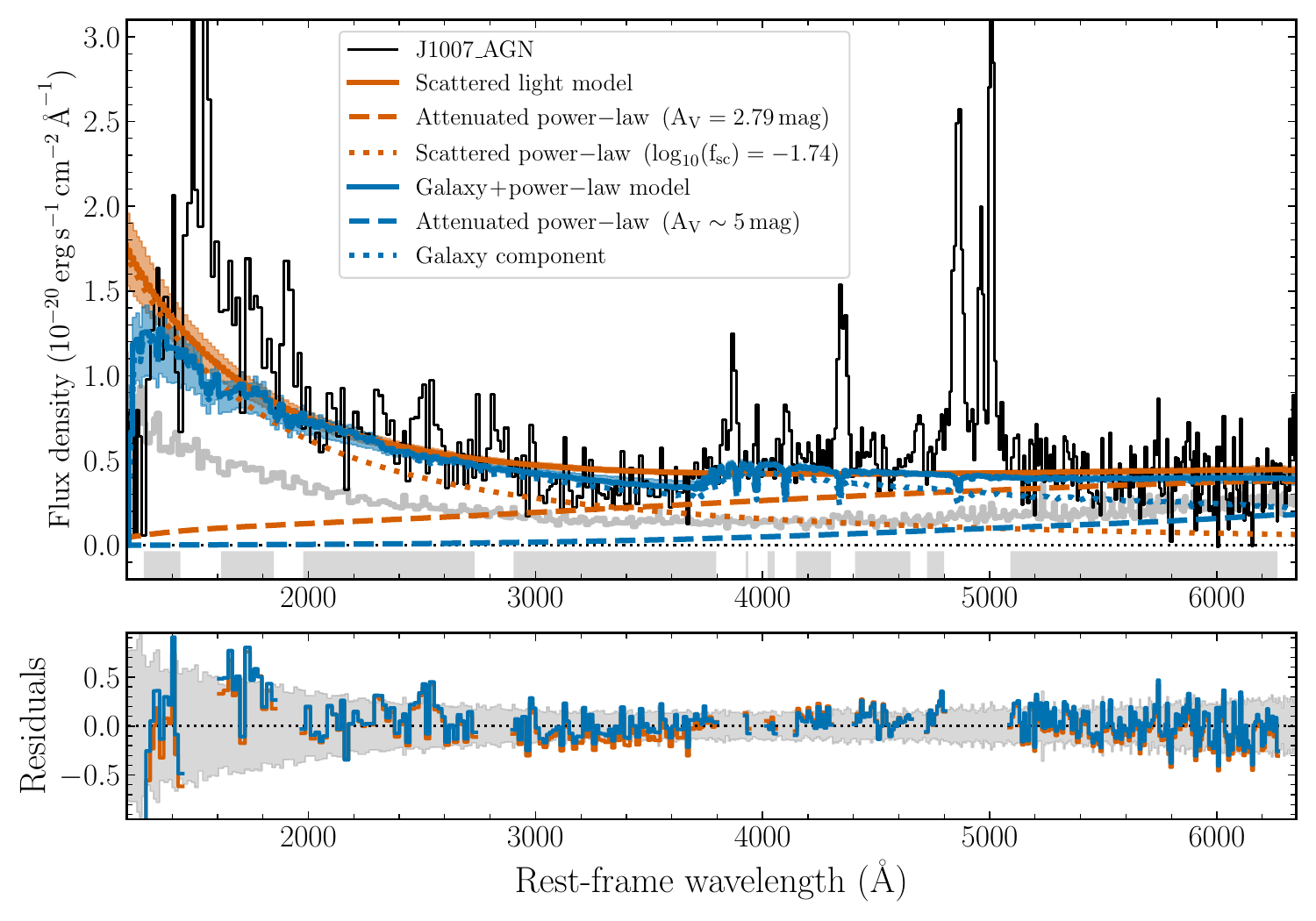}
\caption{\small 
\textbf{Continuum model fits to the \agn\ spectrum.} \textit{Top panel:} The scattered light power-law model (orange) and the galaxy + power-law model (blue). Individual fit components are highlighted with dashed and dotted lines as described in the legend. The model is fit to the \agn\ spectrum (black) in the gray shaded emission-line free regions indicated at the bottom. The 68-percentile posterior of the continuum model is shown in light orange/blue around the median fit.
\textit{Bottom panel:} Residual of the median continuum model fits (orange/blue) contrasted by the 1$\sigma$ flux uncertainties of the \agn\ spectrum (gray regions). 
}
\label{fig:av_fit}
\end{figure}

\begin{figure}[ht!]
\centering
\includegraphics[width=0.92\linewidth]{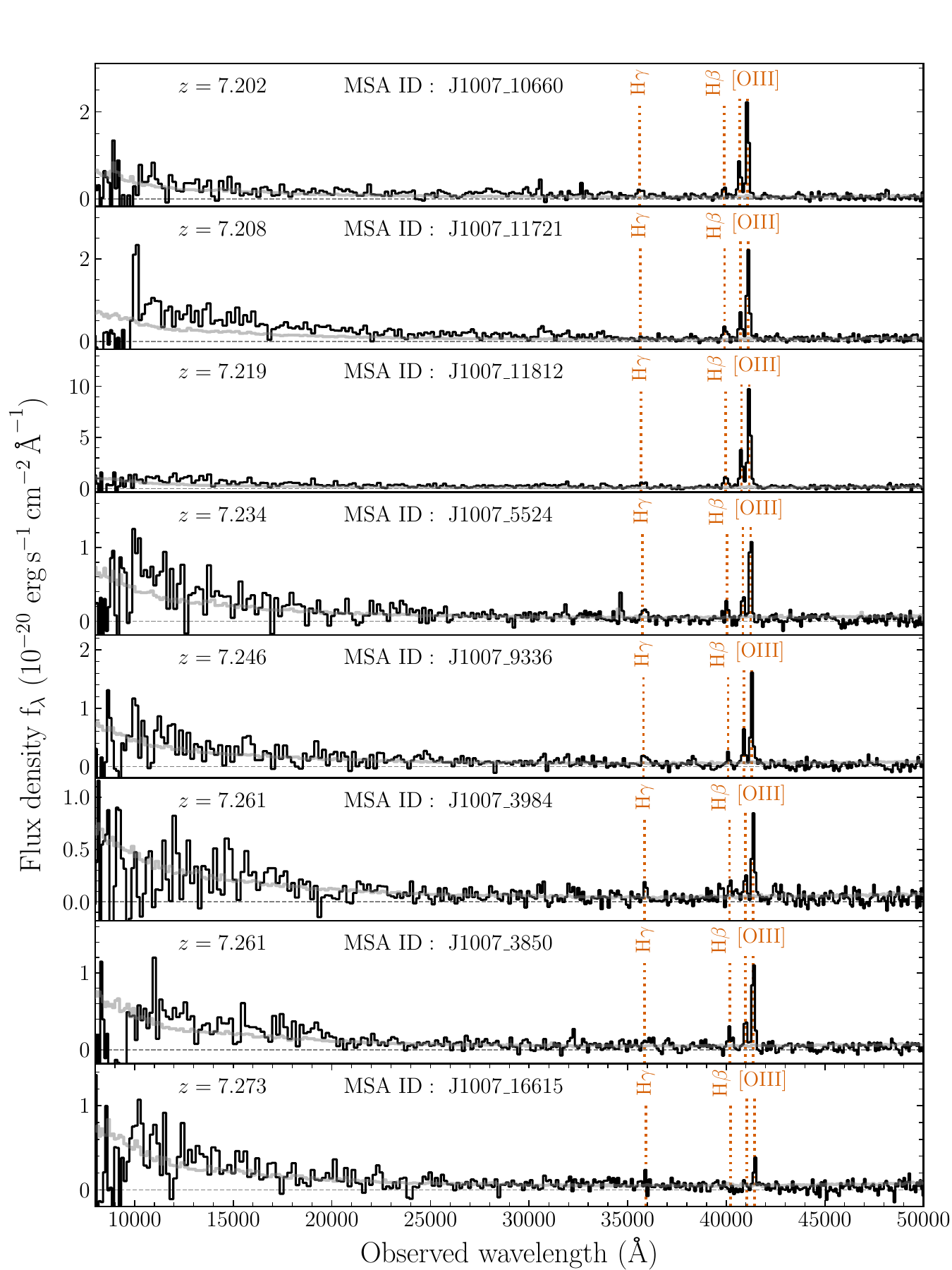}
\caption{\small 
\textbf{Optimally extracted and co-added NIRSpec/MSA spectra of the 8 galaxies within $\mathbf{2500\,\textrm{\bf km}\,\textrm{\bf s}^{-1}}$ along the line of sight relative to the AGN.} We display the spectral flux (black) and its corresponding $1\sigma$ uncertainty (grey). The redshifts are derived from spectral fits to the \oiiibr $\lambda\lambda 5007,4959$ doublet. Notable emission lines are indicated in orange.}
\label{fig:galaxy_discovery}
\end{figure}

\begin{figure}[ht!]
\centering
\includegraphics[width=0.38\textwidth]{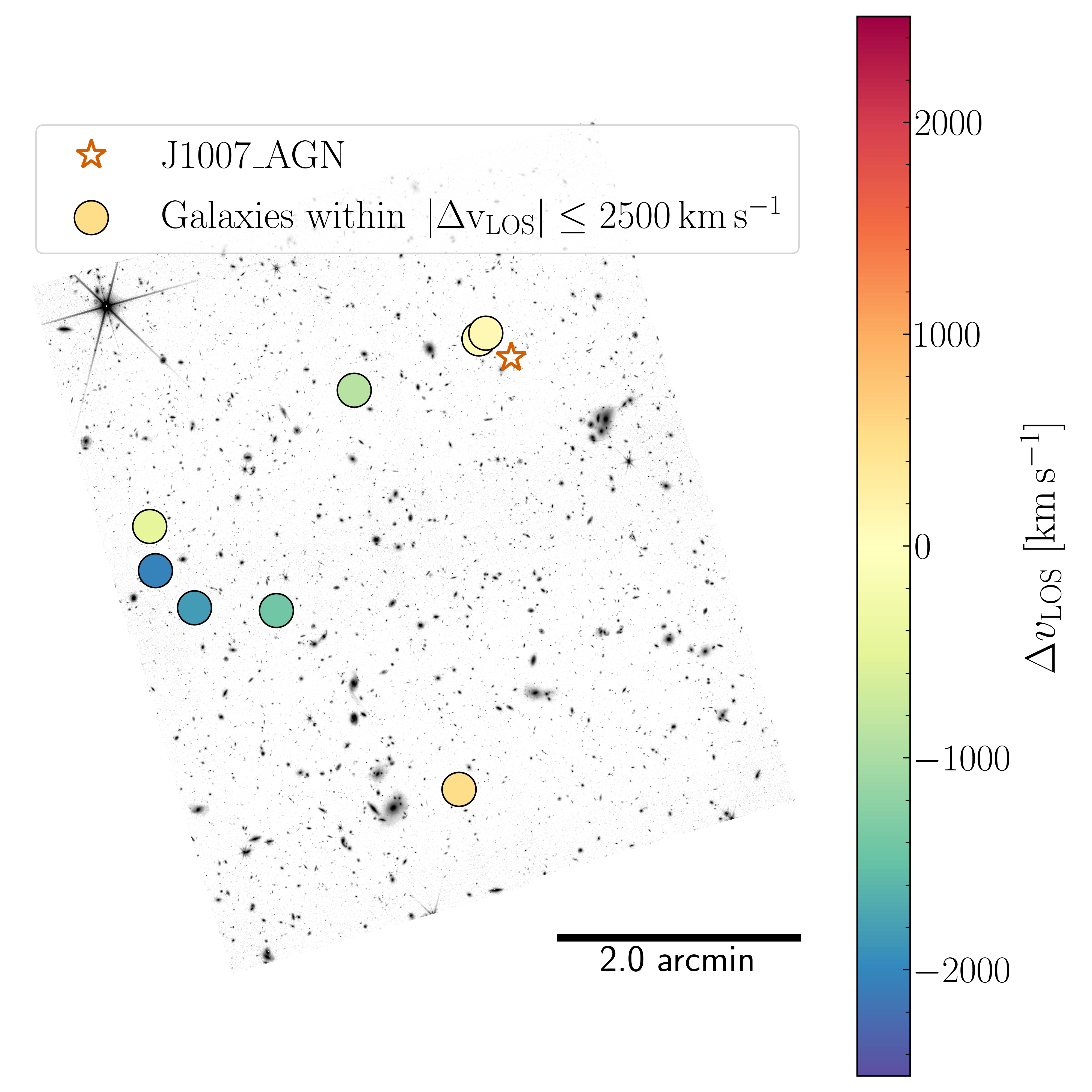}
\includegraphics[width=0.6\textwidth]{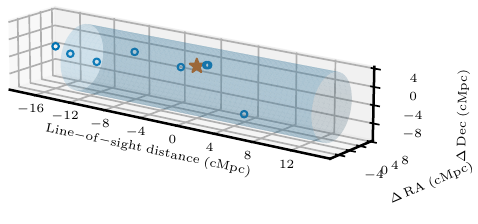}
\caption{\small
\textbf{Distribution of \agn and associated galaxies with $\left| \Delta  v_{\textrm{LOS}} \right| \le 2500\,\textrm{km}\,\textrm{s}^{-1}$ in the field of the sky and visualized in 3D, including the line-of-sight distance.} \textit{Left panel:}
\agn\ (orange star) and galaxy (circles) positions highlighted on the F277W mosaic of the J1007 quasar field. Galaxy positions are coloured by the line-of-sight velocity. 
\textit{Right panel:} A 3D (sky position and line-of-sight distance) representation of \agn\ and its eight nearby galaxies. The maximum volume for the cross-correlation analysis ($\left| \Delta  v_{\textrm{LOS}} \right| \le 1500\,\textrm{km}\,\textrm{s}^{-1}$; $\left| r \right| \le10.9\,\textrm{cMpc}\ $) is highlighted by the blue cylinder, encompassing six of the eight galaxies.
}
\label{fig:galaxy_positions}
\end{figure}


\renewcommand{\tablename}{Extended Data Table}

\setcounter{table}{0}

\begin{table}[h!]
\renewcommand{\arraystretch}{1.3}
\footnotesize 
\centering 
\caption{\textbf{AGN Properties}}
\begin{tabular}{llll} 
\small 
 Property as a function of & $R\approx273$ (default) &  $R\approx278$ & $R\approx268$ \\ 
 resolution at [OIII]\,5007 &  FWHM=$1100\,\textrm{km}\,\textrm{s}^{-1}$ & FWHM=$1080\,\textrm{km}\,\textrm{s}^{-1}$ & FWHM=$1120\,\textrm{km}\,\textrm{s}^{-1}$ \\ 
\hline 
Redshift &  \multicolumn{3}{c}{ $7.2583_{-0.0005}^{+0.0005}$ }  \\ 
$M_{1450}/(\textrm{mag})$ & \multicolumn{3}{c}{$-19.97_{-0.45}^{+0.77}$} \\ 
$\alpha_{\textrm{OPT}}$ & $0.28_{-0.33}^{+0.36}$  & $0.33_{-0.29}^{+0.29}$ & $0.16_{-0.32}^{+0.30}$  \\ 
\hline 
$\textrm{FWHM}_{\textrm{narrow}}/(\textrm{km}\,\textrm{s}^{-1})$ & $431.46_{-82.66}^{+84.63}$  & $422.03_{-76.19}^{+113.49}$ & $449.21_{-68.16}^{+86.96}$  \\ 
$\textrm{FWHM}_{\textrm{broad}}/(\textrm{km}\,\textrm{s}^{-1})$ & $3370.06_{-648.27}^{+1155.54}$  & $3563.62_{-772.50}^{+1174.64}$ & $3535.58_{-766.42}^{+1187.76}$  \\ 
$\textrm{EW}_{\textrm{[OIII]}5007}/(\textrm{\AA})$ & $149.15_{-26.50}^{+24.17}$  & $148.82_{-28.33}^{+27.17}$ & $160.48_{-23.04}^{+25.40}$  \\ 
$\textrm{EW}_{\textrm{[OIII]}4960}/(\textrm{\AA})$ & $62.57_{-13.74}^{+13.23}$  & $61.96_{-14.52}^{+12.79}$ & $68.10_{-13.22}^{+15.84}$  \\ 
$\textrm{EW}_{\textrm{[NeIII]}3869}/(\textrm{\AA})$ & $44.02_{-9.54}^{+11.11}$  & $44.18_{-9.38}^{+10.58}$ & $42.71_{-8.02}^{+9.78}$  \\ 
$\textrm{Flux}_{\textrm{H}\beta, \textrm{broad}}/(10^{-20}\,\textrm{erg}\,\textrm{s}^{-1}\,\textrm{cm}^{-2})$ & $622.95_{-75.32}^{+82.82}$  & $588.29_{-76.70}^{+83.21}$  & $634.66_{-90.63}^{+94.62}$  \\ 
$\textrm{Flux}_{\textrm{H}\beta, \textrm{narrow}}/(10^{-20}\,\textrm{erg}\,\textrm{s}^{-1}\,\textrm{cm}^{-2})$ & $359.09_{-181.97}^{+236.19}$  & $385.14_{-174.04}^{+289.20}$  & $323.66_{-178.48}^{+176.36}$  \\ 
$\textrm{Flux}_{\textrm{H}\gamma, \textrm{broad}}/(10^{-20}\,\textrm{erg}\,\textrm{s}^{-1}\,\textrm{cm}^{-2})$ & $102.03_{-56.89}^{+72.67}$  & $106.03_{-63.78}^{+80.29}$  & $99.01_{-64.73}^{+84.04}$  \\ 
$\textrm{Flux}_{\textrm{H}\gamma, \textrm{narrow}}/(10^{-20}\,\textrm{erg}\,\textrm{s}^{-1}\,\textrm{cm}^{-2})$ & $84.46_{-29.61}^{+29.83}$  & $82.17_{-30.66}^{+29.55}$  & $87.56_{-29.96}^{+31.70}$  \\ 
\hline 
$\log(L_{5100}/( \textrm{erg}\,\textrm{s}^{-1}\,\textrm{\AA}^{-1}))$  & $40.43_{-0.02}^{+0.03}$  & $40.43_{-0.02}^{+0.02}$ & $40.42_{-0.02}^{+0.02}$  \\ 
$\log(L_{\textrm{bol}}/(\textrm{erg}\,\textrm{s}^{-1}))$  & $45.10_{-0.02}^{+0.03}$  & $45.11_{-0.02}^{+0.02}$ & $45.09_{-0.02}^{+0.02}$  \\ 
$\log(L_{\textrm{H}\beta}/(\textrm{erg}\,\textrm{s}^{-1}))$  & $42.78_{-0.07}^{+0.09}$  & $42.78_{-0.07}^{+0.11}$ & $42.77_{-0.06}^{+0.07}$  \\ 
$\log(L_{\textrm{bol, H}\beta}/(\textrm{erg}\,\textrm{s}^{-1}))$  & $45.52_{-0.06}^{+0.08}$  & $45.52_{-0.06}^{+0.09}$ & $45.51_{-0.05}^{+0.06}$  \\ 
$M_{\textrm{BH, GH05, LH}\beta}/(10^{7}\,\textrm{M}_{\odot})$  & $11.52_{-4.63}^{+10.11}$  & $12.69_{-5.13}^{+10.89}$ & $12.20_{-5.16}^{+10.81}$  \\ 
$\lambda_{\textrm{Edd, GH05, LH}\beta}$  & $0.20_{-0.09}^{+0.13}$  & $0.18_{-0.08}^{+0.12}$ & $0.19_{-0.09}^{+0.14}$  \\ 
\hline 
$M_{\textrm{BH, VP06}}/(10^{7}\,\textrm{M}_{\odot})$  & $10.80_{-3.67}^{+8.44}$  & $12.11_{-4.63}^{+9.24}$ & $11.76_{-4.50}^{+8.89}$  \\ 
$\lambda_{\textrm{Edd, VP06}}$  & $0.09_{-0.04}^{+0.05}$  & $0.08_{-0.04}^{+0.05}$ & $0.08_{-0.04}^{+0.05}$  \\ 
$M_{\textrm{BH, GH05}}/(10^{7}\,\textrm{M}_{\odot})$  & $6.11_{-2.06}^{+4.74}$  & $6.86_{-2.62}^{+5.23}$ & $6.62_{-2.52}^{+4.96}$  \\ 
$\lambda_{\textrm{Edd, GH05}}$ & $0.16_{-0.07}^{+0.09}$  & $0.15_{-0.06}^{+0.10}$ & $0.15_{-0.07}^{+0.10}$  \\ 
$M_{\textrm{BH, P12}}/(10^{7}\,\textrm{M}_{\odot})$  & $8.94_{-2.69}^{+5.78}$  & $9.88_{-3.37}^{+6.28}$ & $9.60_{-3.27}^{+6.00}$  \\ 
$\lambda_{\textrm{Edd, P12}}$  & $0.11_{-0.05}^{+0.05}$  & $0.10_{-0.04}^{+0.06}$ & $0.10_{-0.04}^{+0.06}$  \\ 
\hline 
\hline 
\label{tab:spec_info} 
\end{tabular} 
\end{table}

\begin{figure}[ht!]
\centering
\includegraphics[width=0.54\textwidth]{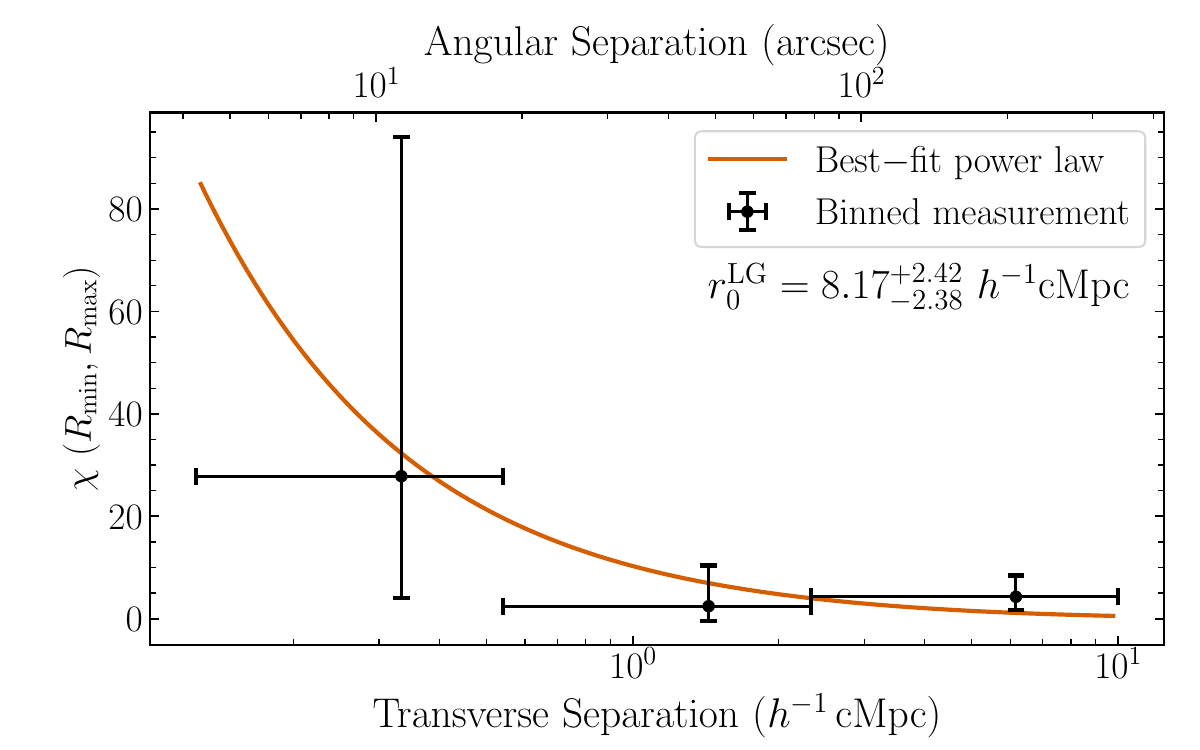}
\includegraphics[width=0.43\textwidth]{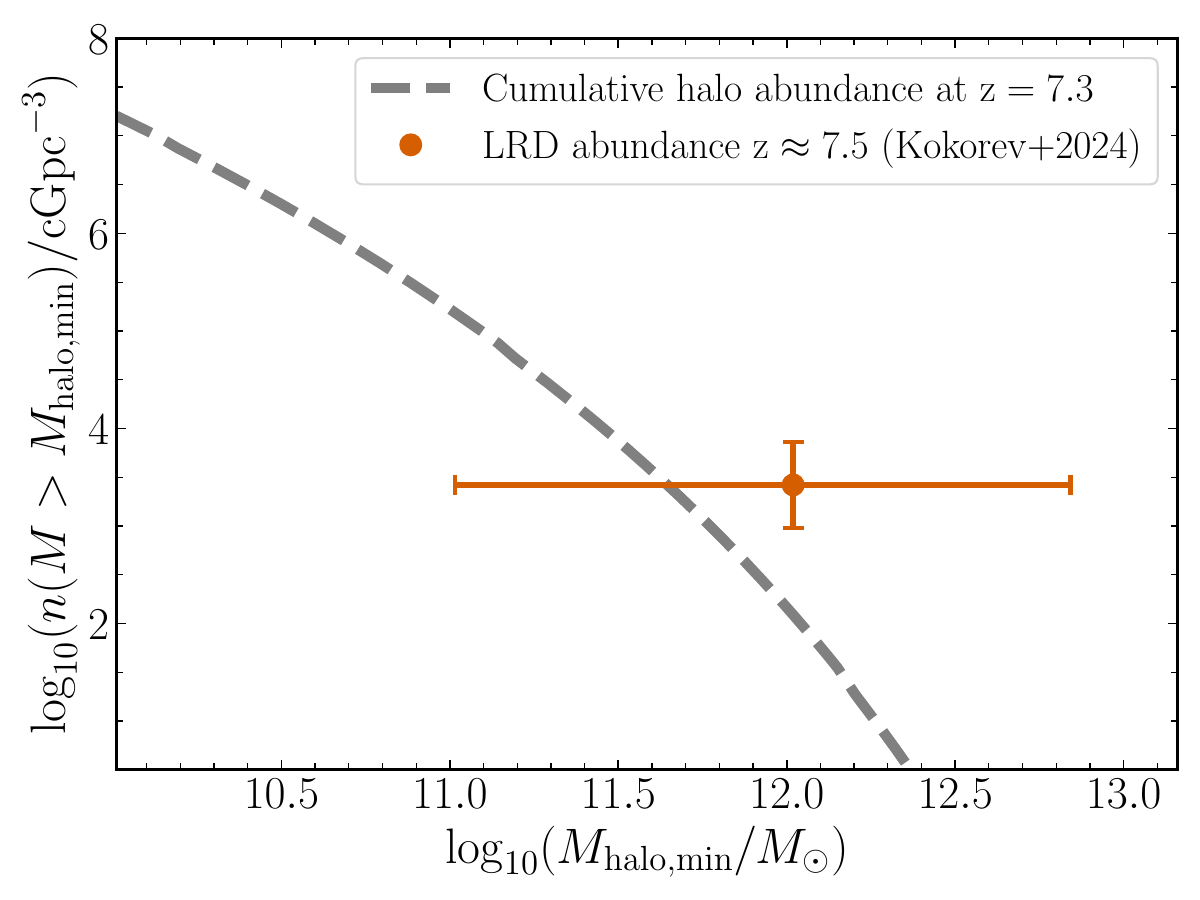}
\caption{\small
\textbf{Clustering measurement and a comparison between the cumulative halo mass function and the LRD abundance at our derived minimum halo mass.} \textit{Left panel:}  Volume-averaged cross-correlation function $\chi$ as a function of transverse separation in three radial bins. We fit the LRD-galaxy two-point correlation function $\xi_{\textrm{LG}}=(r/r_0^{\textrm{LG}})^{-2.0}$ to the 6 galaxies within $\left| \Delta  v_{\textrm{LOS}} \right| \le 1500\,\textrm{km}\,\textrm{s}^{-1}$, and report our best-fit value of the cross-correlation length $r_0^{\textrm{LG}}$.
Uncertainties reflect the confidence interval for a Poisson distribution that corresponds to $1\sigma$ in Gaussian statistics\cite{Gehrels1986}.
\textit{Right panel:} Cumulative halo abundance as a function of minimum halo mass. Adopting a recent measurement for the LRD abundance\cite{Kokorev2024} at $z\approx7.5$, we depict our minium halo mass estimate in orange.
Error bars reflect the $1\sigma$ (16 to 84 percentile) statistical uncertainties. At face value, the combination of LRD abundance and our $M_{\textrm{halo,min}}$ mass range is largely in excess of the cumulative halo abundance at $z=7.3$. 
}
\label{fig:clustering}
\end{figure}

\begin{table}[h!]
\renewcommand{\arraystretch}{1.3}
\footnotesize 
\centering 
\caption{\textbf{Dereddened AGN Properties}}
\begin{tabular}{ll} 
\small 
 Property &  Value \\ 
\hline 
$m_{1450}/(\textrm{mag})$ &$22.67_{-0.49}^{+0.52}$ \\ 
$M_{1450}/(\textrm{mag})$ &$-24.32_{-0.49}^{+0.52}$ \\ 
$\log(L_{5100}/(\textrm{erg}\,\textrm{s}^{-1}\,\textrm{\AA}^{-1}))$ &$41.43_{-0.12}^{+0.11}$ \\ 
$\log(L_{\textrm{bol}}/(\textrm{erg}\,\textrm{s}^{-1}))$ &$46.10_{-0.12}^{+0.11}$ \\ 
\hline 
$\log(L_{\textrm{H}\beta}/(\textrm{erg}\,\textrm{s}^{-1}))$ &$44.05_{-0.14}^{+0.15}$ \\ 
$\log(L_{\textrm{bol, H}\beta}/(\textrm{erg}\,\textrm{s}^{-1}))$ &$46.64_{-0.12}^{+0.13}$ \\ 
$M_{\textrm{BH, GH05, LH}\beta}/(10^{7}\,\textrm{M}_{\odot})$ &$45.14_{-23.20}^{+52.18}$ \\ 
$\lambda_{\textrm{Edd, GH05, LH}\beta}$ &$0.58_{-0.31}^{+0.62}$ \\ 
\hline 
\hline 
\label{tab:dereddened_prop} 
\end{tabular} 
\end{table}

\begin{table}[h!]
\renewcommand{\arraystretch}{1.3}
\footnotesize 
\centering 
\caption{\textbf{Source information}}
\begin{tabular}{llllllll} 
\small 
 Target ID &  R.A. (J2000) & Dec. (J2000) & $z_{\textrm{OIII}}$  & F090W & F115W & F277W & F444W \\ 
   &  \multicolumn{2}{c}{(decimal degrees)} &    & (nJy) & (nJy) & (nJy) & (nJy) \\ 
\hline 
\agn & 151.978226 & 21.283994 & 7.2583 & 4.85 $\pm$ 3.90 & 70.07 $\pm$ 4.65 & 92.00 $\pm$ 2.58 & 421.30 $\pm$ 3.52 \\ 
\hline 
J1007\_10660 & 152.030665 & 21.254673 & 7.2022 & 2.75 $\pm$ 7.70 & 65.71 $\pm$ 8.64 & 82.61 $\pm$ 5.57 & 184.26 $\pm$ 7.25 \\ 
J1007\_11721 & 152.024914 & 21.249581 & 7.2083 & 12.28 $\pm$ 8.89 & 94.01 $\pm$ 8.44 & 113.79 $\pm$ 6.03 & 193.79 $\pm$ 9.03 \\ 
J1007\_11812 & 152.012827 & 21.249211 & 7.2193 & 2.50 $\pm$ 6.87 & 35.67 $\pm$ 6.94 & 53.74 $\pm$ 3.43 & 229.07 $\pm$ 5.38 \\ 
J1007\_5524 & 152.001375 & 21.279459 & 7.2341 & 1.24 $\pm$ 4.64 & 43.39 $\pm$ 4.54 & 46.23 $\pm$ 2.43 & 101.88 $\pm$ 3.52 \\ 
J1007\_9336 & 152.031510 & 21.260753 & 7.2458 & -0.13 $\pm$ 4.42 & 21.59 $\pm$ 4.28 & 32.42 $\pm$ 2.66 & 70.22 $\pm$ 4.00 \\ 
J1007\_3984 & 151.982968 & 21.286523 & 7.2605 & 0.48 $\pm$ 4.62 & 21.68 $\pm$ 4.61 & 15.11 $\pm$ 2.31 & 34.05 $\pm$ 3.34 \\ 
J1007\_3850 & 151.981961 & 21.287311 & 7.2612 & 3.59 $\pm$ 4.15 & 35.25 $\pm$ 4.30 & 26.62 $\pm$ 2.40 & 50.84 $\pm$ 3.16 \\ 
J1007\_16615 & 151.985916 & 21.224606 & 7.2725 & -3.89 $\pm$ 6.73 & 35.31 $\pm$ 6.79 & 34.98 $\pm$ 3.03 & 44.93 $\pm$ 4.88 \\
\hline 
\hline 
\label{tab:source_info} 
\end{tabular} 
Note: In order to make the velocity shifts in Extended Data Table\,\ref{tab:galaxy_info} consistent with the quoted redshift here, we provide a higher accuracy for the redshift than the nominal redshift uncertainty of $\sigma_z \approx 0.001$ from the fit.  
\end{table}

\begin{table}[h!]
\renewcommand{\arraystretch}{1.3}
\footnotesize 
\centering 
\caption{\textbf{Galaxy properties relative to J1007\_AGN}}
\begin{tabular}{ccccccccc} 
\small 
 Target ID & Priority & $\Delta v_{\textrm{LOS}}$ & Angular separation & Angular separation & $M_{\textrm{UV}}$  & $L_{\textrm{[OIII]}5008}$ & $EW_{\textrm{[OIII]}5008}$\\ 
   &  & $(\textrm{km}\,\textrm{s}^{-1})$ & (arcseconds) & $(\textrm{pkpc})$ & (mag) & $(10^{42}\,\textrm{erg}\,\textrm{s}^{-1})$ & $(\textrm{\AA})$ \\ 
\hline 
J1007\_10660 & 1 & $-2038$ & 205.16 & 1049.60 & ${-20.13}_{-0.13}^{+0.15}$& ${2.36}_{-0.08}^{+0.09}$& ${782.26}_{-72.66}^{+71.38}$\\ 
J1007\_11721 & 1 & $-1817$ & 199.70 & 1021.69 & ${-20.51}_{-0.09}^{+0.10}$& ${2.33}_{-0.08}^{+0.08}$& ${676.56}_{-56.38}^{+67.74}$\\ 
J1007\_11812 & 1 & $-1416$ & 170.75 & 873.55 & ${-19.46}_{-0.19}^{+0.24}$& ${10.83}_{-0.25}^{+0.23}$& ${1215.29}_{-82.08}^{+99.38}$\\ 
J1007\_5524 & 1 & $-879$ & 79.35 & 405.96 & ${-19.68}_{-0.11}^{+0.12}$& ${1.26}_{-0.07}^{+0.08}$& ${954.51}_{-169.54}^{+237.18}$\\ 
J1007\_9336 & 1 & $-455$ & 197.37 & 1009.73 & ${-18.93}_{-0.20}^{+0.24}$& ${1.52}_{-0.08}^{+0.08}$& ${1217.91}_{-215.92}^{+201.29}$\\ 
J1007\_3984 & 1 & 80 & 18.33 & 93.76 & ${-18.94}_{-0.21}^{+0.26}$& ${0.88}_{-0.06}^{+0.07}$& ${470.24}_{-65.06}^{+81.53}$\\ 
J1007\_3850 & 1 & 107 & 17.31 & 88.55 & ${-19.46}_{-0.12}^{+0.14}$& ${1.24}_{-0.07}^{+0.07}$& ${607.25}_{-80.22}^{+112.45}$\\ 
J1007\_16615 & 2 & 516 & 215.35 & 1101.72 & ${-19.47}_{-0.19}^{+0.23}$& ${0.26}_{-0.05}^{+0.06}$& ${125.90}_{-30.53}^{+30.96}$\\ 
\hline 
\hline 
\label{tab:galaxy_info} 
\end{tabular} 
Note: Given an accuracy on the emission line redshifts of $\sigma_z \approx 0.001$ the velocity along the line of sight has an uncertainty of $\approx 40\,\textrm{km}\,\textrm{s}^{-1}$. 
\end{table}

\begin{table}[h!]
\footnotesize 
\centering 
\renewcommand{\arraystretch}{1.3}
\caption{\textbf{AGN galaxy cross-correlation results}}
\begin{tabular}{ll|llll} 
\small 
 $R_{\textrm{min}}$ & $R_{\textrm{max}}$ & $\langle LG\rangle$  & $\langle LR\rangle$ & $\chi_{LG}$ & $\chi_{LG,\textrm{corr}}$\\ 
 $(\textrm{cMpc}\ h^{-1})$  & $(\textrm{cMpc}\ h^{-1})$ & & & &  \\ 
\hline 
\multicolumn{6}{l}{$\left|\Delta v_{\textrm{LOS}}\right| \le 1500\,\textrm{km}\,\textrm{s}^{-1}$ } \\ \hline 
0.13 & 0.54 & 1 &  0.03 & $27.82_{-23.84}^{+66.28}$ & $27.82_{-23.84}^{+66.28}$ \\ 
0.54 & 2.33 & 1 &  0.35 & $1.86_{-2.37}^{+6.58}$ & $2.45_{-2.85}^{+7.93}$ \\ 
2.33 & 10.0 & 4 &  1.86 & $1.15_{-1.03}^{+1.70}$ & $4.28_{-2.53}^{+4.17}$ \\ 
\hline 
\multicolumn{6}{r}{$r_0^{\textrm{LG}}=6.36_{-2.01}^{+2.09}\,\textrm{cMpc}\ h^{-1}$\ \ ; $r_{0,\textrm{corr}}^{\textrm{LG}}=8.17_{-2.38}^{+2.42}\,\textrm{cMpc}\ h^{-1}$ } \\ 
\hline 
\multicolumn{6}{l}{$\left|\Delta v_{\textrm{LOS}}\right| \le 2500\,\textrm{km}\,\textrm{s}^{-1}$ } \\ \hline 
0.13 & 0.54 & 1 &  0.06 & $16.29_{-14.31}^{+39.77}$ & $16.29_{-14.31}^{+39.77}$ \\ 
0.54 & 2.33 & 1 &  0.58 & $0.72_{-1.42}^{+3.95}$ & $1.07_{-1.71}^{+4.76}$ \\ 
2.33 & 10.0 & 6 &  3.10 & $0.94_{-0.77}^{+1.16}$ & $3.30_{-1.71}^{+2.57}$ \\ 
\hline 
\multicolumn{6}{r}{$r_0^{\textrm{LG}}=6.22_{-2.06}^{+2.13}\,\textrm{cMpc}\ h^{-1}$\ \ ; $r_{0,\textrm{corr}}^{\textrm{LG}}=8.28_{-2.47}^{+2.45}\,\textrm{cMpc}\ h^{-1}$ } \\ 
\hline 
\hline 
\label{tab:cross_corr} 
\end{tabular} 
\end{table}

\clearpage

\bibliography{all_new}

\end{document}